\documentclass[12pt,a4paper]{article}
\usepackage{amssymb, amsmath,framed,upgreek}
\usepackage{mathrsfs}
\usepackage{dsfont}
\usepackage[colorlinks]{hyperref}
\usepackage{tikz} 
\usepackage{subcaption}
\newcommand{\rem}[1]{}
\newcommand{\de}{{\rm d}}

\newcommand{\bx}{\boldsymbol{x}}
\newcommand{\bsigma}{\boldsymbol\sigma}

\newcommand{\bZ}{\,\widehat{\!\boldsymbol{\cal Z}\,}\!}

\newcommand{\hmH}{{\widehat{\mathcal{H}}}}

\newcommand{\bp}{\boldsymbol{p}}
\newcommand{\bX}{{\mathbf{X}}}
\newcommand{\bbX}{{\boldsymbol{\mathcal{X}}}}

\newcommand{\bz}{{\mathbf{z}}}

\newcommand{\vtb}{{\boldsymbol{\vartheta}}}

\newcommand{\beq}{\begin{equation}}
\newcommand{\eeq}{\end{equation}}

\newcommand{\comment}[1]{\vspace{1 mm}\par
\marginpar{\large\underline{}}\noindent
\framebox{\begin{minipage}[c]{0.95 \textwidth}
\rm\tt #1 \end{minipage}}\vspace{2 mm}\par}

\newtheorem{remark}{Remark}

\renewcommand{\contentsname}{}

\begin{document}

\title{Heisenberg dynamics of mixed quantum-classical systems\footnote{Contribution to  the Focus Point ``\href{https://link.springer.com/journal/13360/topicalCollection/AC_db80eeeca18f684acf512bff52efa4d2}{Mathematics and Physics at the Quantum-Classical Interface}'' in the European Physical Journal Plus.}}
\author{David Mart\'inez-Crespo$^{1,2}$, Cesare Tronci$^3$
\\\vspace{-.1cm}
\footnotesize
\it $^1$Departamento de F\'isica Te\'orica, Universidad de Zaragoza,  Zaragoza, Spain
\\\vspace{-.1cm}
\footnotesize
\it $^2$Centro de Astropartículas y Física de Altas Energías, Universidad de Zaragoza,  Zaragoza, Spain
\\\vspace{-.1cm}
\footnotesize
\it $^3$School of Mathematics and Physics, University of Surrey, Guildford, United Kingdom}
\date{\vspace{-.5cm}}

\maketitle

\begin{abstract}
We consider the dynamics of interacting quantum and classical systems in the Heisenberg representation. Unlike the usual construction in standard quantum mechanics, mixed quantum-classical systems involve the interplay of unitary operators acting on the quantum observables and the Lagrangian trajectories sweeping the classical degrees of freedom. This interplay reflects  an intricate structure which is made particularly challenging by the backreaction excerpted on the classical trajectories by the quantum degrees of freedom. While the backreaction is underestimated in the common  Ehrenfest model, more recent methodologies succeed in capturing  this important effect by resorting to Koopman wavefunctions in classical mechanics. Luckily, both Ehrenfest and Koopman models enjoy a variational framework which is exploited here to unfold the geometric structure underlying quantum-classical coupling. A special role is played by the action of the diffeomorphic Lagrangian paths on a non-Abelian pure-gauge potential  which comprises statistical correlations. After presenting the treatment in the simple case of Ehrenfest dynamics, we move on to the Koopman model and present the role of the backreaction terms therein. Finally, we compare both models in the context of pure-dephasing systems.
\vspace{-1cm}
\end{abstract}
{
\contentsname
\tableofcontents
}

\section{Introduction and background}

The formulation of dynamical models for the coupled dynamics of  quantum and classical systems represents an open problem in different contexts, ranging from molecular chemistry \cite{CrBa18} to solid-state physics \cite{MaHuHe19}, and the theory of gravity \cite{AlBoClMa24,GiGrSc23}, to name a few. In the first two cases, the computational challenges posed by fully quantum many-particle simulations motivate the search for approximate descriptions in which parts of the system are treated classically while the remainder is left fully quantum. In the context of gravity, the difficulties in achieving a quantum theory of gravity may pave the way to approximate formulations in which space-time is treated classically, while matter retains its fully quantum nature.

\subsection{Models of hybrid quantum-classical dynamics}
Mean-field approaches to hybrid quantum-classical dynamics have long been known. 
In spite of their intrinsic simplicity, the lack of statistical correlations therein makes these theories obsolete and insufficient in most situations. Several remarkable attempts to retain quantum-classical correlations are found in the literature although most of these  lead to the violation of important consistency properties such as Heisenberg's uncertainty principle. For example, this is the case for the so-called \emph{quantum-classical Liouville equation} \cite{Aleksandrov,Gerasimenko,Kapral}, which currently represents a standard approach in chemical physics. Other computational schemes, such as the celebrated \emph{surface-hopping} \cite{Bondarenko,Tully90}, suffer from the same drawback.

Among the various approaches made available over the years \cite{Anderson,Traschen,Diosi,Hall,JaSu10,Kapral,PrKi,Sudarshan}, the \emph{Ehrenfest model} appears to be the only one  retaining an entire list of consistency properties \cite{Traschen,GBTr22}, including the uncertainty principle. This model provides the dynamics of a distribution-valued density matrix $\widehat{P}(q,p,t)$, where $(q,p)$ are classical coordinates, such that $f(q,p)=\operatorname{Tr}\widehat{P}(q,p)$ and $\hat\varrho=\int\widehat{P}(q,p)\,\de q\de p$ are the classical Liouville density and  the quantum density matrix, respectively  \cite{alonsoEffectiveNonlinear2023}. Upon introducing the conditional density matrix $\hat\rho(q,p)=\widehat{P}(q,p)/f(q,p)$, we may write the Ehrenfest model as follows:
\beq\label{Ehrenfest}
\frac{\partial f}{\partial t}
+\operatorname{div}(f\langle\bX_{\widehat{H}}\rangle)=0
\,,\qquad\qquad
i\hbar\frac{\partial \hat\rho}{\partial t}+i\hbar\langle\bX_{\widehat{H}}\rangle\cdot\nabla\hat\rho=[\widehat{H},\hat\rho],
\eeq
where $\widehat{H}(q,p)$ is a Hamiltonian Hermitian operator depending on the classical coordinates and
$\bX_{\widehat{H}}=(\partial_p\widehat{H},-\partial_q\widehat{H})$ is the mixed quantum-classical Hamiltonian vector field. Also, we have introduced the notation $\langle\widehat{A}(q,p)\rangle=\operatorname{Tr}(\hat\rho(q,p)\widehat{A}(q,p))$. We observe that the second equation in \eqref{Ehrenfest} is independent of the classical Liouville density, which in turn is only enslaved to the quantum motion. This decoupling of the latter from the classical Liouville dynamics represents a major limitation for this model, which is indeed known to fail in accurately capturing correlation effects such as quantum decoherence. The latter can be expressed in terms of the quantum purity $\|\hat\varrho\|^2=\operatorname{Tr}\hat\varrho^2$ and is a crucial property of quantum systems in interaction. Some attempts are found in the literature to restore decoherence in Ehrenfest dynamics by \emph{ad-hoc} methods \cite{AkLoPr14}.  Thus, despite its physical consistency \cite{Alonso}, the Ehrenfest model is insufficient in most relevant contexts. 

In recent years, our team has proposed an alternative model of mixed quantum-classical dynamics that not only retains physical consistency,  but also reproduces decoherence levels with high accuracy. Successfully tested in \cite{BaBeGBTr24} for several problems in physics and chemistry, the model is based on the theory of Koopman wavefuctions \cite{Koopman,Mauro} and builds up on early work by George Sudarshan \cite{Marmo,Sudarshan}. Koopman wavefunctions are  square-integrable phase-space functions $\chi(q,p,t)$ such that the density $f=|\chi|^2$ satisfies the classical Liouville equation. In \cite{Sudarshan}, Sudarshan proposed to consider mixed quantum-classical dynamics on the tensor-product Hilbert space of Koopman and Schr\"odinger wavefunctions. For example, in the case of infinite-dimensional quantum systems, this leads to writing $\widehat{P}(q,p,x,x')=\Upsilon(q,p,x)\Upsilon(q,p,x')^*$. Here, $\Upsilon(q,p,x)$ is a hybrid quantum-classical wavefunction, so that $f(q,p)=\int|\Upsilon(q,p,x)|^2\,\de x$ and we recall $\hat\rho=\widehat{P}/f$. While the construction of a dynamical model on such a tensor-product Hilbert space is a very intuitive idea, its {realization} is far from obvious and only recently was our team able to present the resulting equations of motion. Their formulation went through several stages \cite{BoGBTr19,GBTr22,GBTr20} and has  been reviewed in \cite{TrGB23}. Importantly, the quantum-classical model based on Koopman wavefunctions is Hamiltonian and, similarly to the Ehrenfest equations \eqref{Ehrenfest}, enjoys a variational formulation in terms of Hamilton's action principle. In particular, the model appears as an $\hbar$-correction of Ehrenfrest dynamics, where the correcting terms make the corresponding  equations of motion rather formidable. The latter are as follows:
\beq
\partial_t f+\operatorname{div}(f\boldsymbol{X})=0
\,,\qquad\ 
i\hbar(\partial_t\hat\rho+\boldsymbol{X}\cdot\nabla\hat\rho)=\big[\,\widehat{\!\mathscr{ H}},\hat\rho\big],
\label{HybEq1}
\eeq
with
\beq
\boldsymbol{X}=
\langle\bX_{\widehat{H}}\rangle+\frac\hbar{2f}\operatorname{Tr}\!\big( 
\bX_{{\widehat{H}}}\cdot\nabla \widehat{\boldsymbol\Sigma}
-
\widehat{\boldsymbol\Sigma}\cdot\nabla\bX_{{\widehat{H}}}
\big),
\qquad\quad
\widehat{\boldsymbol\Sigma}=i f[\hat\rho,\bX_{\hat\rho}],
\label{HybEq2}
\eeq
and
\begin{align}%\nonumber
\,\widehat{\!\mathscr{ H}}
%&={\widehat{H}+i\hbar\big[\nabla{\hat\rho}+\hat\rho\nabla{\ln}\sqrt{f},\bX_{{\widehat{H}}}\big]}\\
&= \widehat{H}+i\hbar\Big(\{\hat\rho,{\widehat{H}}\}+\{\widehat{H},\hat\rho\}-\frac{1}{2}[\{ {\ln} f,\widehat{H}\},\hat\rho] \Big),
\label{HybEq3}
\end{align}  
Despite their intimidating appearance, it was pointed out in \cite{BaBeGBTr24,TrGB23} that these equations do not involve gradients of order higher than two. This is in contrast, for example, with fully quantum approaches based on quantum hydrodynamics that contain third-order derivatives \cite{Madelung}. Nevertheless, expanding the different terms in the phase-space vector field $\boldsymbol{X}$ makes this system hardly approachable by conventional methods. Luckily, however, the equations \eqref{HybEq1}-\eqref{HybEq3} enjoy a variational formulation which has so far served as the main operational platform for devising various extensions and reduced models, from quantum-classical spin dynamics \cite{GBTr23} to fluid closures \cite{GBTr-fluid} and  numerical particle schemes \cite{BaBeGBTr24}. {Recently, we also identified an entropy functional for this system \cite{TrMCGB}}.

\subsection{Mixed quantum-classical Heisenberg dynamics}
So far, most works on mixed quantum-classical models have only dealt with the dynamics of states, i.e. the quantum and the classical states in interaction. A few authors, however, have posed the more fundamental question of the interaction of quantum and classical \emph{observables}, as given by operator-valued functions on phase-space \cite{Anderson,Gerasimenko2,PeTe,Sahoo,boutheliermadre2023}. In particular, special emphasis has been put on how the quantum backreaction affects the classical motion. Also, questions emerge about the algebraic nature of these observables: given that classical and quantum observables comprise a Lie-algebra structure under the Poisson bracket and the commutator, respectively, is there an analogue Lie-algebra structure for quantum-classical observables? Similar questions were recently investigated in \cite{GBTr20} by resorting to the geometry of hybrid quantum-classical wavefunctions. {While this paper shows how the evolution of hybrid observables is generated by a Lie-algebra action underlying a group structure, the observables themselves fail to inherit such a Lie-algebra structure}.
%In this paper, we show that such a Lie-algebra structure may generally be absent.

 As customary in standard quantum mechanics, this type of questions prompts the need to develop a Heisenberg representation of mixed quantum-classical dynamics. Other than allowing the direct evaluation of expectation values, the Heisenberg picture is useful in many situations. Examples are given by the rotating-wave approximation in Rabi dynamics and other situations in scattering theory where  Heisenberg dynamics appears as an intermediate description in Dirac's interaction picture. The duality between the Schr\"odinger and the  Heisenberg representation in quantum mechanics has a direct counterpart in classical mechanics, and in particular in the dynamics of continuum media. On the one hand, the  quantum Schr\"odinger picture corresponds to the classical \emph{Eulerian description}, which amounts to describing the continuum flow by looking at it from the inertial laboratory frame. In terms of phase-space dynamics, this picture corresponds to the dynamics of the classical state, as given by the classical Liouville equation $\partial_tf+\{f,H\}=0$. On the other hand, the  quantum Heisenberg picture corresponds to the classical \emph{convective description}, which is widely common in elasticity \cite{SiMaKr88} and lets the observer move together with the medium. In terms of phase-space dynamics, this picture corresponds to the dynamics of classical observables, as given by $\partial_t\mathcal{O}=\{\mathcal{O},H\}$ for any function $\mathcal{O}(q,p,t)$. A third picture -- the \emph{Lagrangian description} -- is also available in classical fluid dynamics and consists in studying the evolution of the entire Lagrangian fluid trajectory, as expressed by a diffeomorphism of the spatial region occupied by the fluid. In the case of phase-space dynamics, if $\boldsymbol\eta(q,p,t)$ is a Lagrangian path, i.e. a time-dependent diffeomorphism of phase-space, its dynamics is given by $\dot{\boldsymbol\eta}=\bX_{H}\circ\boldsymbol\eta$, where $\circ$ denotes composition of mappings and we recall that $\bX_{H}$ denotes the Hamiltonian vector field. In quantum mechanics, this picture corresponds to considering the dynamics of the unitary propagator, that is $i\hbar\dot{U}=\widehat{H}U$. 
 \begin{figure}[h!]\center
\includegraphics[scale=.5925]{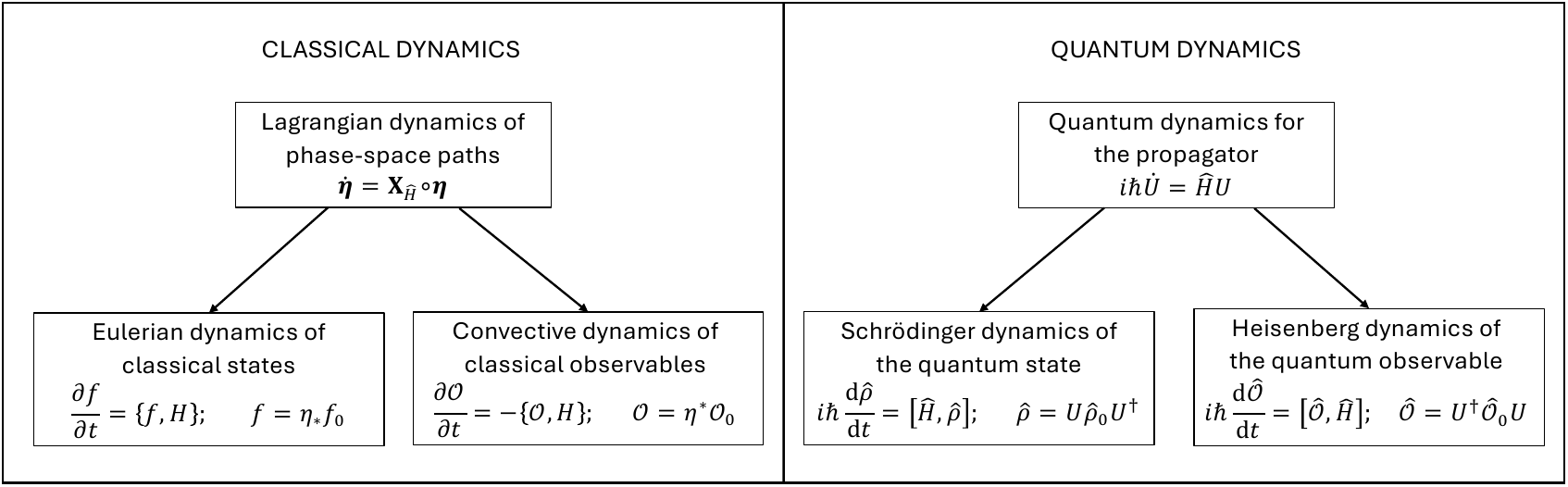}
\end{figure}
%%%%%%%%%%%%%%%%

%%%%%%%%%%%%%%%%

In this paper, we will consider the Heisenberg representation for both the Ehrenfest equations \eqref{Ehrenfest} and the Koopman model in \eqref{HybEq1}-\eqref{HybEq3}. In this context, the formulation of Heisenberg dynamics is made rather subtle by the time-dependent coupling between the Lagrangian phase-space trajectories advancing the classical motion and the unitary operators governing the quantum dynamics. Indeed, as shown by the material derivatives $\partial_t+\langle\bX_{\widehat{H}}\rangle\cdot\nabla$ and $\partial_t+\boldsymbol{X}\cdot\nabla$ appearing in the second equation of \eqref{Ehrenfest} and \eqref{HybEq1}, respectively, the quantum dynamics is expressed in the frame moving with the phase-space vector field advancing the classical flow, that is $\langle\bX_{\widehat{H}}\rangle$ and $\boldsymbol{X}$, respectively. This crucial aspect of mixed quantum-classical dynamics makes the search for a Heisenberg representation far from easy. Luckily, once again, the variational formulation provides invaluable help and eventually leads to the final answer. As we will see, the action principle underlying quantum-classical dynamics also leads to a rich geometric construction involving the dynamics of non-Abelian gauge connections on principal bundles. 

Before proceeding, however, we will review the variational structure on quantum-classical dynamics in the conventional (Schr\"odinger) picture. Then, the Heisenberg dynamics will be obtained by operating on the action principle and in particular on the group actions involved therein.

\paragraph{Plan of the paper.} Following this introductory discussion, in Section \ref{sec:QCVP} we review  the Eulerian formulation of the Koopman hybrid model \eqref{HybEq1}-\eqref{HybEq3} based on Euler-Poincaré reduction by symmetry. Particular emphasis will be {placed} on the appearance of the Wilczek-Zee connection in the Hamiltonian functional. As we will see, this connection {form} plays a prominent role in quantum-classical Heisenberg dynamics.  In Section \ref{sec:Ehrenfest} we deal with the Heisenberg picture of the Ehrenfest model. In spite of its simplicity, this model retains  
all the essential geometric properties that will also be present in the more advanced context of Koopman quantum-classical dynamics.
%every ingredient needed to derive the equations of motion from the Euler-Poincaré variational principle in the convective frame. 
%In particular, we will show how the convective counterpart of the Wilczek-Zee connection plays a predominant role in the Heisenberg derivation. 
We will discuss the purely classical and quantum specializations of the description along as well as comment on the conservation laws in the model. In Section \ref{sec:Koopman} we use similar techniques to deal with the Koopman model. The classical part of the dynamics, i.e. the generator of diffeomorphic phase-space paths, deserves special attention and thus we devote Section \ref{sec:KoopmanVectorField} to its study. In particular we develop a new algebraic tool that, inside a pairing, acts over operators as a Lie derivative with respect to an operator-valued vector field. This {valuable} tool allows us to unveil the geometric nature underlying the quantum back-reaction in the Koopman hybrid model. We use this approach in \ref{sec:HeisembergFull} to present the equations of motion of an arbitrary hybrid operator undergoing hybrid Koopman dynamics. Conservation laws are also discussed again in this more involved setting. In Section \ref{sec:pureDephasing} we exemplify the Heisenberg equations of motion of the Koopman model for the simple case of a pure-dephasing Hamiltonian. Finally we conclude in Section \ref{sec:conclusions} with a general discussion and we provide an outline of possible future directions.

\subsection{Quantum-classical variational setting and  gauge connections}
 \label{sec:QCVP}
As anticipated, this section sets up the ground by reviewing the geometric setting for  quantum-classical variational models in the Eulerian Schr\"odinger picture.
The variational principle $\delta\int_{t_1}^{t_2}\ell\,\de t=0$ underlying both the Ehrenfest equations \eqref{Ehrenfest} and the Koopman model in \eqref{HybEq1}-\eqref{HybEq3} is based on Euler-Poincar\'e reduction theory \cite{HoMaRa98}, which relates the dynamical quantities to the group actions determining the time evolution. In particular, mixed quantum-classical dynamics involves a Lagrangian of the type
\beq
\ell(\boldsymbol{X},\hat\xi,f,\hat\rho)=\int\! f\big(\boldsymbol{\cal A}\cdot\boldsymbol{X}+\langle\hat\rho,i\hbar\hat\xi\rangle\big)\,\de^2z-h(f,\hat\rho).
\label{MQCLagr}
\eeq
Here, $\bz=(q,p)$, so that $\de^2z=\de q\,\de p$, and $\boldsymbol{\cal A}\cdot\de\bz=p\de q$ is the canonical one-form. Also, $\langle A,B\rangle=\operatorname{Re}\operatorname{Tr}(A^\dagger B)$ is the real-valued pairing on quantum operators. If $\eta(t)\in\operatorname{Diff}(T^*Q)$ is a time-dependent diffeomorphism on the phase-space $T^*Q=\Bbb{R}^2$,  and $U(\bz,t)$ is a time-dependent unitary operator depending parametrically on the phase-space coordinates, we have the following definitions:
\beq
f:=\eta_*f_0
,\qquad\quad
\hat\rho:=(U\hat\rho_0U^\dagger)\circ\boldsymbol\eta^{-1}
,\qquad\quad
\boldsymbol{X}:=\dot{\boldsymbol\eta}\circ\boldsymbol\eta^{-1}
,\qquad\quad
\hat\xi:=\dot{U}U^\dagger\circ\boldsymbol\eta^{-1}.
\label{LtoE}
\eeq
Here, $\circ$ denotes composition of functions and $\eta_*f_0=(f_0/\det\nabla\boldsymbol\eta)\circ\boldsymbol\eta^{-1}$ is the  push-forward of the volume form $f_0\,\de^2z$, which is then Lie-transported by the Lagrangian trajectory $\eta$. In particular, the second relation in \eqref{LtoE} shows how the unitary quantum evolution is expressed in the frame moving with the classical trajectory $\eta$. The vector field ${X}=\boldsymbol{X}(\bz,t)\cdot\nabla$ and the skew-Hermitian operator $\hat\xi(\bz,t)$ determine the generators of the classical and quantum flows, respectively. In summary, the density matrix  $\hat\rho$ evolves on orbits of the semidirect-product group $\operatorname{Diff}(T^*Q)\,\circledS\,{\cal F}(T^*Q,\mathscr{U}(\mathscr{H}_\text{\tiny\sf Q}))$, where ${\cal F}(T^*Q,\mathscr{U}(\mathscr{H}_\text{\tiny\sf Q}))$ denotes the space of smooth functions from the phase-space $T^*Q=\Bbb{R}^2$ to the group $\mathscr{U}(\mathscr{H}_\text{\tiny\sf Q})$ of unitary operators on the quantum Hilbert space $\mathscr{H}_\text{\tiny\sf Q}$. Also, $({X},\hat\xi)\in\mathfrak{X}(T^*Q)\,\circledS\,{\cal F}(T^*Q,\mathfrak{u}(\mathscr{H}_\text{\tiny\sf Q}))$, where $\mathfrak{X}(T^*Q)$ is the Lie algebra of phase-space vector fields and $\mathfrak{u}(\mathscr{H}_\text{\tiny\sf Q})$ is the Lie algebra of skew-Hermitian operators on $\mathscr{H}_\text{\tiny\sf Q}$. Notice that, while here we deal with the simplest case $Q=\Bbb{R}$, the present treatment can be easily extended to consider smooth configuration manifolds.

As customary in Euler-Poincar\'e reduction \cite{HoMaRa98}, the definitions  \eqref{LtoE} lead to the following constrained variations:
\[
\delta f=-\pounds_{ Y}f
,\qquad\qquad\ 
\delta{X}=\partial_t{ Y}+\pounds_{ X}{ Y},
\]
and
\[
\delta \hat\rho = [\hat\varsigma, \hat\rho] - \pounds_{ Y}\hat\rho
,\qquad\qquad\ 
\delta \hat\xi = \partial _t \hat\varsigma + [ \hat\varsigma , \hat\xi  ] + \pounds_{ X}  \hat\varsigma   -  \pounds_{ Y}   \hat\xi,
\]
where $\boldsymbol{Y}=\delta\boldsymbol\eta\circ\boldsymbol\eta^{-1}$ and $\hat\varsigma=(\delta{U}U^\dagger)\circ\boldsymbol\eta^{-1}$ are both arbitrary quantities vanishing at the endpoints.
Here, we have introduced the Lie derivative operator $\pounds$ and we recall the Jacobi-Lie coordinate formula for the Lie derivative of vector fields, that is $\pounds_{\boldsymbol{X}}\boldsymbol{Y }=\boldsymbol{X}\cdot\nabla\boldsymbol{Y}-\boldsymbol{Y}\cdot\nabla\boldsymbol{X}$ and the notation is such that $\pounds_{{X}}{Y }=\pounds_{\boldsymbol{X}}\boldsymbol{Y }\cdot\nabla$. Thus, upon writing $X=\boldsymbol{X}\cdot\nabla$ and $Y=\boldsymbol{Y}\cdot\nabla$, in local coordinates we have
\[
\delta f=-\operatorname{div}(f\boldsymbol{Y})
,\qquad\qquad\ 
\delta\boldsymbol{X}=\partial_t\boldsymbol{Y}+\boldsymbol{\cal X}\cdot\nabla\boldsymbol{Y}-\boldsymbol{Y}\cdot\nabla\boldsymbol{\cal X},
\]
and
\[
\delta \hat\rho = [\hat\varsigma, \hat\rho] - \boldsymbol{Y}\cdot\nabla\hat\rho
,\qquad\qquad\ 
\delta \hat\xi = \partial _t \hat\varsigma + [ \hat\varsigma , \hat\xi  ] + \boldsymbol{ X}  \cdot\nabla  \hat\varsigma   -  \boldsymbol{Y}\cdot\nabla   \hat\xi.
\]
Then, taking variations of the Lagrangian \eqref{MQCLagr} under the action principle ${\delta\int_{t_1}^{t_2}\ell\,\de t=0}$ yields
\beq
\boldsymbol{ X}=\bX_{\textstyle\frac{\delta h}{
\delta f}}-\frac1f\bigg\langle \frac{\delta h}{\delta \hat\rho},\bX_{\hat\rho}\bigg\rangle
%+\frac1D\bigg\langle{\cal P}\bigg| \bX_{\textstyle\frac{\delta h}{\delta {\cal P}}}\bigg\rangle
,\qquad\qquad 
i\hbar[   \hat\xi,\hat\rho]=\frac1f\bigg[\frac{\delta h}{
\delta \hat\rho},\hat\rho\bigg],
\label{EPeqns}
\eeq
which  need to be replaced in the equations
\[
\partial_t f+\operatorname{div}(f\boldsymbol{ X})=0\,,\qquad\qquad
\partial_t \hat\rho + \boldsymbol{ X}\cdot\nabla\hat\rho = [\hat\xi, \hat\rho] \,,
\]
obtained by taking the time derivative of the first two equations in \eqref{LtoE}. 

At this point, the Ehrenfest model is obtained by inserting the Hamiltonian functional
\[
h(f,\hat\rho)=\int\! f\langle\hat\rho,\widehat{H}\rangle\,\de^2z
\]
in \eqref{MQCLagr}, while the equations \eqref{HybEq1}-\eqref{HybEq3} follow, after a few rearrangements, from
\beq
h(f,\hat\rho)=\int \!f\big\langle\hat\rho,\widehat{H}+{i\hbar}\{\hat\rho,\widehat{H}\}\big\rangle\,\de^2z,
\label{KoopHam}
\eeq
where $\{\cdot,\cdot\}$ is the canonical Poisson bracket. We observe that the last term above contains first-order gradients accounting for the inhomogeneities associated to quantum-classical correlations. 
\begin{remark}[Analogies to spin-orbit coupling]
The last term in the Hamiltonian functional \eqref{KoopHam} has been found \cite{GBTr-fluid} to have analogies to spin-orbit coupling  (SOC) in semirelativistic mechanics \cite{Baym,Thomas}. To see this, it is useful to suitably project so that 
$\langle\hat\rho,i\{\hat\rho,\widehat{H}\}\rangle=\langle i[\hat\rho,\nabla\hat\rho],\bX_{\widehat{H}}\rangle/2$. Then, one shows that an analogous term appears in the energy associated to spin-orbit coupling. In that context, the  Hamiltonian operator is $\widehat{H}=m^{-1}|\hat\bp|^2/2+{V(\hat{\bx})}+\widehat{\cal H}_{SOC}$, where $ \hat{\bp}=-i\hbar\nabla$  and 
\[
\widehat{H}_{SOC}=-\frac\hbar{4 m^2 \mathsf{c}^2}\,\widehat{\bsigma}\times\nabla V\cdot\hat{\boldsymbol{p}}.
\]
As usual, the  Pauli equation $i\hbar\partial_t\Psi=\widehat{H}\Psi$ is obtained from the Dirac-Frenkel action principle $\delta\int_{t_1}^{t_2}\!\int\operatorname{Re}(i\hbar\Psi^\dagger\partial_t\Psi-\Psi^\dagger\widehat{H}\Psi)\,\de t=0$, where $\Psi$ is  a Pauli spinor  such that ${\int\Psi^\dagger\Psi\,\de^3x=1}$. Following \cite{BBirula}, we write $\Psi(\bx)=\chi(\bx)\phi(\bx)$. Here,  $\chi$ is a spatial wavefunction and $\phi(\bx)$ is a scalar field of unit vectors, that is $\|\phi(\bx)\|=1$. Then, the Dirac-Frenkel  action principle becomes $\delta\int_{t_1}^{t_2}\!\int\operatorname{Re}
\big(i\hbar\chi^*\partial_t\chi+|\chi|^2\langle\phi,i\hbar\partial_t\phi\rangle - \langle\chi\phi,\widehat{H}(\chi\phi)\rangle\big)\de^3 x\de t=0$, and the SOC integrand  reads
\[
\langle\chi\phi,\widehat{H}_{SOC}(\chi\phi)\rangle=\frac\hbar{4 m^2 {c}^2}\langle\widehat{\bX}_{SOC}\rangle\cdot\operatorname{Re}(\chi^*\hat{\boldsymbol{p}}\chi+|\chi|^2\mathbf{A})+\frac\hbar{4 m^2 {c}^2}|\chi|^2\langle i[\hat\varrho,\nabla\hat\varrho],\widehat{\bX}_{SOC}
\rangle
,
\]
where $\hat\varrho=\phi\phi^\dagger$, $\widehat{\bX}_{SOC}=\nabla V\times\widehat{\bsigma}$ is the SOC vecvtor field, and $\mathbf{A}=\langle\phi,-i\hbar\nabla\phi\rangle$ is the Berry connection. We observe the direct analogy between the last term above and the backreaction term in  \eqref{KoopHam}, after suitably projecting  therein. This intriguing analogy will be developed elsewhere.
\end{remark}

The gradients appearing in the last term of \eqref{KoopHam} may be expressed in terms of a non-Abelian gauge connection first appeared in the work by Wilczek and Zee \cite{WiZe84} and playing a predominant role in the main part of this paper. To see how this connection emerges, we can use the second {equation} in \eqref{LtoE} to compute $\nabla\hat\rho=[\hat\rho,\widehat{\boldsymbol\lambda}]$, where 
\beq\label{EulConn}
\widehat{\boldsymbol\lambda}=\eta_*\big(U\widehat{\boldsymbol\lambda}_0U^\dagger-(\nabla U)U^\dagger\big).
\eeq
Here, $\widehat{\boldsymbol\lambda}_0$ is a fixed operator-valued differential one-form  such that $\nabla\hat\rho_0=[\hat\rho_0,\widehat{\boldsymbol\lambda}_0]$, while  $\eta_*$ is the push-forward. Then, the gauge connection $\widehat{\lambda}=\widehat{\boldsymbol\lambda}\cdot\de\bz$ appears in the Hamiltonian \eqref{KoopHam}, which is rewritten as
\beq
h(f,\hat\rho,\widehat{\boldsymbol\lambda})=\int \!\Big(f\langle\hat\rho,\widehat{H}\rangle+\frac12\langle[\hat\rho,\widehat{\boldsymbol\lambda}],i\hbar[\bX_{\widehat{H}},\hat\rho]\rangle\Big)\,\de^2z
\label{KoopHam2}.
\eeq
Taking derivatives of the definition \eqref{EulConn} yields the coordinate-free relations
\begin{equation*}
\partial_t\widehat{\lambda}=-\pounds_{{ X}}\widehat{\lambda}-\de^{\widehat{\lambda}}{\hat\xi}
\,,\qquad\qquad
\delta\widehat{\lambda}=-\pounds_{{Y}}\widehat{\lambda}-\de^{\widehat{\lambda}}{\hat\varsigma},
\end{equation*}
where we have introduced the exterior differential $\de$ so that, in index notation, $(\de\widehat{\boldsymbol\lambda})_{jk}=\partial_j\widehat{\lambda}_k-\partial_k\widehat{\lambda}_j$. In addition, $\de^{\widehat{\lambda}}\,\cdot=\de\cdot+[\,{\widehat{\lambda}},\cdot\,]$ is the covariant differential.
In local coordinates, $\widehat{\lambda}=\widehat{\boldsymbol\lambda}\cdot\de\bz$ and $X=\boldsymbol{ X}\cdot\nabla$, so that
\begin{align*}
\partial_t\widehat{\boldsymbol\lambda}=-{\boldsymbol{ X}}\cdot\nabla\widehat{\boldsymbol\lambda}-\nabla\boldsymbol{ X}\cdot\widehat{\boldsymbol\lambda}-\nabla{\hat\xi}-[{\widehat{\boldsymbol\lambda}},\hat\xi]
\,,\qquad\qquad
\delta\widehat{\boldsymbol\lambda}=-{\boldsymbol{ Y}}\cdot\nabla\widehat{\boldsymbol\lambda}-\nabla\boldsymbol{ Y}\cdot\widehat{\boldsymbol\lambda}-\nabla{\hat\varsigma}-[{\widehat{\boldsymbol\lambda}},\hat\varsigma].
\end{align*}
 We empahsize that, as already noted in e.g. \cite{FoTr24}, the parallel transport relation $\nabla\hat\rho_0=[\hat\rho_0,\widehat{\boldsymbol\lambda}_0]$, that is $\de^{\widehat{\lambda}_0}\hat\rho_0=0$, implies that the initial connection  $\widehat{\boldsymbol\lambda}_0$ is flat, that is $\de^{\widehat{\lambda}_0}\widehat{\lambda}_0=0$, thereby implying $\de^{\widehat{\lambda}}\widehat{\lambda}=0$ at all times. 
For example, the second Ehrenfest equation in \eqref{Ehrenfest} may be expressed as $
i\hbar{\partial_t \hat\rho}=[\widehat{H}-i\hbar\langle\bX_{\widehat{H}}\rangle\cdot{\widehat{\boldsymbol\lambda}},\hat\rho]$. Analogously, the model equations \eqref{HybEq1}-\eqref{HybEq3} may be rewritten in terms of the Wilczek-Zee connection by making repeated use of the parallel transport relation $\nabla\hat\rho=[\hat\rho,\widehat{\boldsymbol\lambda}]$.
Instead of performing this replacement, here we will move on to presenting the variational approach to the Heisenberg representation by first considering the Ehrenfest model \eqref{Ehrenfest}. As we will see, the Heisenberg dynamics of the Wilczek-Zee connection will play a dominant role in the present study.

\section{Heisenberg picture of hybrid Ehrenfest dynamics\label{sec:Ehrenfest}}
In this section, we will perform various transformations in such a way to express the Lagrangian \eqref{MQCLagr} in the Heisenberg representation. As customary, we will freeze the \emph{state variables} $f$ and $\hat\rho$, so that the usual duality relation $\int A\,\eta_*(B\de^2z)=\int (B\,\eta^* A)\de^2z$ leads to
\[
\int\! f\big(\boldsymbol{\cal A}\cdot\boldsymbol{ X}+\langle\hat\rho,i\hbar\hat\xi-\widehat{H}\rangle\big)\,\de^2z=\int\! f_0\big(\boldsymbol{\Theta}\cdot\boldsymbol{\cal X}+\langle\hat\rho_0,i\hbar\widehat\varpi-\widehat{\cal H}\rangle\big)\,\de^2z_0
\]
where we recall ${\cal A}=\boldsymbol{\cal A}\cdot\de\bz=p\de q$ and define
\beq
{\Theta}=\eta^*{\cal A}
,\qquad\quad
\widehat{\cal H}={U}^\dagger(\widehat{H}\circ\boldsymbol\eta){U}
,\qquad\quad
{\cal X}:=\eta^*(\dot{\boldsymbol\eta}\circ\boldsymbol\eta^{-1})
,\qquad\quad
\widehat\varpi:={U}^\dagger\dot{U}.
\label{LtoE-H1}
\eeq
In order to make the treatment as explicit as possible, here we have changed the integration labels by replacing the Eulerian coordinates $\bz=(q,p)$ with their convective counterpart $\bz_0=(q_0,p_0)$. For example, the relation $f=\eta_* f_0$ in \eqref{LtoE} reads explicitly $f(\bz,t)=\int \!f_0(\bz_0)\,\delta(\bz-\boldsymbol\eta(\bz_0,t))\,\de^2 z_0$.

We observe that the Heisenberg Hamiltonian does not evolve on a group orbit because the map $\widehat{H}\mapsto{U}^\dagger(\eta^*\widehat{H}){U}$ does not correspond to a group action. In order to restore the information on group orbits, here we will express the unitary operator $U$ in \eqref{LtoE} in the classical convective  frame by writing $U={\cal U}\circ\boldsymbol\eta$, without loss of generality. {Here, the initial conditions are such that $U|_{t=0}=\boldsymbol\eta|_{t=0}=\boldsymbol{1}$. Then,  also ${\cal U}|_{t=0}=\boldsymbol{1}$ and we obtain $\dot{U}=\partial_t{\cal U}\circ\boldsymbol\eta+{\boldsymbol{\cal X}}\cdot\nabla({\cal U}\circ\boldsymbol\eta)$. B}y a slight abuse of notation, the  forth equation in \eqref{LtoE-H1} becomes
\begin{align}\nonumber
\widehat{\cal H}=({\cal U}^\dagger\widehat{H}{\cal U})\circ\boldsymbol\eta
,\qquad\quad
\widehat\varpi&=({\cal U}^\dagger\dot{\cal U})\circ\boldsymbol\eta+\boldsymbol{\cal X}\cdot\eta^*({\cal U}^\dagger\nabla{\cal U})
\\
&=:\hat\zeta+\boldsymbol{\cal X}\cdot\widehat{\boldsymbol{\gamma}}.
%\hat\xi^{\sf H}+\boldsymbol{\cal X}^{\sf H}\cdot\widehat{\boldsymbol{\gamma}}
\label{mario}
\end{align}
Notice the appearance of the Heisenberg counterpart $\widehat{{\gamma}}=\widehat{\boldsymbol{\gamma}}\cdot\de\bz$ of the Wilczek-Zee connection $\widehat{{\lambda}}$, which appeared previously in the Schr\"odinger picture. This type of gauge connection is now required by the  construction, which otherwise would destroy the group orbit structure as mentioned before. In particular, we notice that the connection $\widehat{\boldsymbol{\gamma}}$ is a \emph{pure-gauge connection}, in the sense that $\widehat{\boldsymbol{\gamma}}|_{t=0}=0$ and $\de^{\widehat{{\gamma}}}\widehat{{\gamma}}=0$.
%Also, the Heisenberg Hamiltonian evolves according to $\widehat{\cal H}=({\cal U}^\dagger\widehat{H}{\cal U})\circ\boldsymbol\eta$. For the sake of generality, we will consider the original Schr\"odinger Hamiltonian $\widehat{H}$ as a matrix analytic function of a certain array of observables, that is operator-valued functions $\widehat{\boldsymbol{O}}=(\widehat{O}_1,\widehat{O}_2,\dots)$, so that we have
%\[
%\widehat{\cal H}=\widehat{H}(\widehat{\boldsymbol{\cal O}})
%\qquad\text{with}\qquad
%\widehat{\boldsymbol{\cal O}}=({\cal U}^\dagger\widehat{\boldsymbol{O}}{\cal U})\circ\boldsymbol\eta
%\]
With these definitions, the Heisenberg Lagrangian reads
\[
\ell_{\sf H}(\boldsymbol{\cal X},\boldsymbol{\Theta},\hat\zeta,\widehat{\boldsymbol{\gamma}}, \widehat{\cal H})=\!\int\!\! f_0\big(\boldsymbol{\Theta}\cdot\boldsymbol{\cal X}+\big\langle \hat\rho_0,i\hbar(\hat\zeta+\boldsymbol{\cal X}\cdot\widehat{\boldsymbol{\gamma}})-\widehat{\cal H}\big\rangle\big)\de^2z_0.
%-h(\widehat{\boldsymbol{\cal O}})
%\quad\text{with}\quad
%h(\widehat{\boldsymbol{\cal O}})=\!\int\!\!f_0\langle\hat\rho_0,\widehat{H}(\widehat{\boldsymbol{\cal O}})\rangle\de^2z.
\]
We observe that the dynamics has now moved from the state variables $f$ and $\hat\rho$ to the Hamiltonian operator $\widehat{\cal H}$ as well as the auxiliary quantities $\Theta=\boldsymbol{\Theta}\cdot\de\bz$ and $\widehat{{\gamma}}=\widehat{\boldsymbol{\gamma}}\cdot\de\bz$. As we will see, {the presence of the Lie-transported  canonical one-form $\Theta=\eta^*{\cal A}$} means that the symplectic properties of the classical phase-space are now evolved in time.
By taking the relevant derivatives in \eqref{LtoE-H1}-\eqref{mario}, we find the auxiliary equations 
\begin{equation}
\label{eq:equationConvective}
\frac{\partial \widehat{\cal H}}{\partial t}+[\hat\zeta,\widehat{\cal H}]=\pounds_{{\cal X}}\widehat{\cal H}
\,,\qquad
\frac{\partial {{\Theta}}}{\partial t}=\pounds_{{\cal X}} {{\Theta}}
\,,\qquad
\frac{\partial \widehat{{\gamma}}}{\partial t}=\pounds_{\cal X}\, {\widehat{{\gamma}}}+\de^{\widehat{{\gamma}}}\hat\zeta,
\end{equation}
along with the variations
\begin{equation}
\label{eq:variationConvective}
\delta\widehat{\cal H}+[\hat\nu,\widehat{\cal H}]=\pounds_{{\cal Y}}\widehat{\cal H}
\,,\qquad
\delta{{\Theta}}=\pounds_{{\cal Y}} {{\Theta}}
\,,\qquad
\delta \widehat{{\gamma}}=\pounds_{{\cal Y}} {\widehat{{\gamma}}}+\de^{\widehat{{\gamma}}}\hat\nu
\,,
\end{equation}
where $\boldsymbol{\cal Y}=\eta^*(\delta\boldsymbol\eta\circ\boldsymbol\eta^{-1})$ and $\hat\nu=(\delta{\cal U}{\cal U}^\dagger)\circ\boldsymbol\eta^{-1}$ are both arbitrary quantities vanishing at the endpoints. We observe that, upon using $\de^{\widehat{{\gamma}}}\widehat{{\gamma}}=0$, the last equations in \eqref{eq:equationConvective} and \eqref{eq:variationConvective} may also be written as
\[
\frac{\partial \widehat{\boldsymbol{\gamma}}}{\partial t}=\nabla^{\widehat{{\gamma}}}(\hat\zeta+\bbX\cdot\widehat{\boldsymbol{\gamma}})
\,,\qquad\quad
\delta \widehat{\boldsymbol{\gamma}}=\nabla^{\widehat{{\gamma}}}(\hat\nu+\boldsymbol{\cal Y}\cdot\widehat{\boldsymbol{\gamma}}),
\] 
respectively. Here, the notation $\nabla^{\widehat{{\gamma}}\,}\cdot=\nabla\cdot\,+\,[\widehat{\boldsymbol{\gamma}},\cdot\,]$ is used whenever we work in local coordinates.
\begin{remark}[Evolution of the connection form]
{We remark that, while setting $\widehat{{\gamma}}_0=0$ is allowed by a convenient global trivialization, the latter is  unavailable in the  case of a nontrivial  $\mathscr{U}(\mathscr{H}_\text{\tiny\sf Q})$-bundle over $T^*Q$. In this more general case, one is forced to retain the arbitrary initial condition $\widehat{{\gamma}}_0$ in the treatment.
%, so that a convenient choice is made available by the particular set of local coordinates. 
In this case, the overall evolution of the connection form reads $\widehat{{\gamma}}_0\mapsto\eta^*({\cal U}^\dagger(\nabla+\widehat{{\gamma}}_0){\cal U})=\widehat{{\gamma}}(t)$. 
Since this evolution law identifies an (affine) action of the semidirect-product group $\operatorname{Diff}(T^*Q)\,\circledS\,{\cal F}(T^*Q,\mathscr{U}(\mathscr{H}_\text{\tiny\sf Q}))$, the connection evolves on orbits thereof.  This type of evolution for connection forms is especially common in  hydrodynamic models of complex fluids and we refer to \cite{GBRa,GBRaTr,Holm02} for more details. For the sake of simplicity, here we will continue to consider the case of a trivial principal bundle and set $\widehat{{\gamma}}_0=0$ in the remainder of this paper.}
\end{remark}
\noindent
More importantly, the variations in \eqref{eq:variationConvective} are accompanied by the relations
\[
\delta{\cal X}=\partial_t{\cal Y}+\pounds_{\cal Y}{\cal X}
\,,\qquad\qquad
\delta \hat\zeta = \partial _t \hat\nu - [ \hat\nu , \hat\zeta  ] +  \pounds_{\cal Y}   \hat\zeta- \pounds_{\cal X}  \hat\nu
\]
which follow from the third definition in \eqref{LtoE-H1} and the defining relation $\hat\zeta:=({\cal U}^\dagger\dot{\cal U})\circ\boldsymbol\eta$.  In local coordinates, we have
\beq
\delta\boldsymbol{\cal X}=\partial_t\boldsymbol{\cal Y}+\boldsymbol{\cal Y}\cdot\nabla\boldsymbol{\cal X}-\boldsymbol{\cal X}\cdot\nabla\boldsymbol{\cal Y}
,\qquad\qquad\ 
\delta \hat\zeta = \partial _t \hat\nu - [ \hat\nu , \hat\zeta  ] +  \boldsymbol{\cal Y}\cdot\nabla   \hat\zeta- \boldsymbol{\cal X}  \cdot\nabla  \hat\nu.
\eeq

Then, the variations $\hat\nu=(\delta{\cal U}{\cal U}^\dagger)\circ\boldsymbol\eta^{-1}$ yield
\beq
\frac{\partial}{\partial t}\frac{\delta \ell_{\sf H}}{\delta \hat\zeta}+\left[\hat\zeta,\frac{\delta \ell_{\sf H}}{\delta \hat\zeta}\right]=\operatorname{div}\!\left(\frac{\delta \ell_{\sf H}}{\delta \hat\zeta}\boldsymbol{\cal X}\right) 
- \operatorname{div}\!\left(\frac{\delta \ell_{\sf H}}{\delta \widehat{\boldsymbol{\gamma}}}\right)
-
\left[\widehat{\boldsymbol{\gamma}},\frac{\delta \ell_{\sf H}}{\delta \widehat{\boldsymbol{\gamma}}}\right]
+
\left[\hmH,\frac{\delta \ell_{\sf H}}{\delta \hmH}\right]
,
\label{eq:quantehr}
\eeq
so that, upon computing 
\[
        \frac{\delta \ell_{\sf H}}{\delta
        \hat\zeta}= -i\hbar f_0\hat\rho_0
        \,,\qquad
        \frac{\delta \ell_{\sf H}}{\delta
        \hmH}= -f_0\hat\rho_0
        \,,\qquad
        \frac{\delta \ell_{\sf H}}{\delta
        \widehat{\boldsymbol{\gamma}}}= -i\hbar f_0\hat\rho_0\boldsymbol{\cal X},
\]
we obtain
\beq\label{infgen}
[\hat\rho_0,i\hbar(\hat\zeta+\boldsymbol{\cal X}\cdot\widehat{\boldsymbol{\gamma}})-\hmH]=0
\implies
\hat\zeta=-\boldsymbol{\cal X}\cdot\widehat{\boldsymbol{\gamma}}-i\hbar^{-1}\hmH+\hat\zeta_0
,\qquad\text{with}\qquad
[\hat\rho_0,\hat\zeta_0]=0.
\eeq
Notice that, in equation \eqref{eq:quantehr}, we denoted $[\widehat{\boldsymbol{\gamma}},{\delta \ell_{\sf H}}/{\delta \widehat{\boldsymbol{\gamma}}}]=[\widehat{{\gamma}}_k,{\delta \ell_{\sf H}}/{\delta \widehat{{\gamma}}}_k]$. For the sake of compactness, this notation will be used throughout the paper wherever possible.

\begin{remark}[Gauge fixing in Heisenberg dynamics]\label{rem1}
The arbitrary skew-Hermitian operator $\hat\zeta_0$ commuting with the initial state $\hat\rho_0$ may appear as a new unexpected object in the present construction of Heisenberg dynamics. However, this arbitrariness is unavoidable and far from accidental. This may be easily seen by considering the second evolution law  in \eqref{LtoE}, that is ${\hat\rho=(U\hat\rho_0U^\dagger)\circ\boldsymbol\eta^{-1}}$. It is clear that the unitary operator $U$ is only defined up to right-multiplication by any unitary operator $U_0$ preserving the initial state $\hat\rho_0$, that is $U_0\hat\rho_0U_0^\dagger=\hat\rho_0$. In other words, the propagator $U$ is only defined up to elements of the isotropy group $G_0=\{U_0\in{\cal F}(T^*Q,\mathscr{U}(\mathscr{H}_\text{\tiny\sf Q})) \,|\,U_0\hat\rho_0U_0^\dagger=\hat\rho_0\}$. This corresponds to a non-Abelian gauge freedom also appearing in the standard quantum treatment of Heisenberg dynamics, as already discussed in \cite{BLTr15}. Accordingly, the infinitesimal generator $\hat\zeta$ of the quantum motion is defined up to elements of the isotropy subalgebra $\mathfrak{g}_0=\{\hat\zeta_0\in{\cal F}(T^*Q,\mathfrak{u}(\mathscr{H}_\text{\tiny\sf Q})) \,|\,[\hat\rho_0,\hat\zeta_0]=0\}$. Since this skew-hermitian operator $\hat\zeta_0$ is an arbitrary gauge choice, one may safely set $\hat\zeta_0=0$ in the final equations. This is the gauge fixing performed throughout this paper.

\end{remark}
In addition, the variations ${\cal Y}=\eta^*(\delta\boldsymbol\eta\circ\boldsymbol\eta^{-1})$  give 
\begin{align}
\frac{\partial}{\partial t}\frac{\delta \ell_{\sf H}}{\delta
\boldsymbol{\cal X}}
=&\, 
\pounds_{\boldsymbol{\cal X}}\,\frac{\delta \ell_{\sf H}}{\delta \boldsymbol{\cal X}}
+
\left\langle\frac{\delta \ell_{\sf H}}{\delta
{\hmH}},\nabla{\hmH}\right\rangle+\left\langle\frac{\delta \ell_{\sf H}}{\delta
\hat\zeta},\nabla{\hat\zeta}\right\rangle
+
\nabla\boldsymbol\Theta\cdot\frac{\delta \ell_{\sf H}}{\delta
\boldsymbol\Theta}
-
\boldsymbol\Theta\operatorname{div}\frac{\delta \ell_{\sf H}}{\delta
\boldsymbol\Theta}
-\frac{\delta \ell_{\sf H}}{\delta
\boldsymbol\Theta}\cdot\nabla\boldsymbol\Theta \nonumber\\
&
+\Big\langle\nabla\widehat{\boldsymbol{\gamma}},\frac{\delta \ell_{\sf H}}{\delta
\widehat{\boldsymbol{\gamma}}}\Big\rangle
-
\Big\langle\operatorname{div}\frac{\delta \ell_{\sf H}}{\delta
\widehat{\boldsymbol{\gamma}}},\widehat{\boldsymbol{\gamma}}\Big\rangle
-
\Big\langle\frac{\delta \ell_{\sf H}}{\delta
\widehat{\boldsymbol{\gamma}}}\cdot\nabla,\widehat{\boldsymbol{\gamma}}\Big\rangle
\label{eq:ehrconv1}
\end{align}
After a few rearrangements using the functional derivatives
\[        
\frac{\delta \ell_{\sf H}}{\delta
        \boldsymbol{\cal X}}= f_0\boldsymbol{\boldsymbol\Theta} + f_0\langle \hat\rho_0,i\hbar\widehat{\boldsymbol{\gamma}} \rangle
        \,,\qquad
        \frac{\delta \ell_{\sf H}}{\delta
        \boldsymbol\Theta}= f_0\boldsymbol{\cal X},
\]
we obtain
\begin{align}
\nonumber
\frac{\partial}{\partial t}\big(\boldsymbol{\Theta}+\langle \hat\rho_0,i\hbar \widehat{\boldsymbol{\gamma}} \rangle\big)
=&\, 
\nabla(\boldsymbol{\cal X}\cdot\boldsymbol{\Theta})
+
\big\langle \hat\rho_0,\nabla(i\hbar \boldsymbol{\cal X}\cdot\widehat{\boldsymbol{\gamma}}
+
i\hbar\hat\zeta
-
\mathcal{H})
\big\rangle
\\
=&\, 
\nabla(\boldsymbol{\cal X}\cdot\boldsymbol{\Theta})
\label{eq:consv1}
\end{align}
where the last equality follows by choosing the gauge $\hat\zeta_0=0$ in \eqref{infgen}. As mentioned in Remark \ref{rem1}, this is the gauge adopted in the remainder of this paper.

\begin{remark}[Role of the Berry connection]
Upon writing $\rho_0=\psi_0\psi_0^\dagger$, with $\psi_0=\eta^*(U^\dagger\psi)$, and introducing the Eulerian Berry connection ${\mathcal{A}}_B=\langle\psi,-i\hbar \de\psi\rangle$ (in the Schr\"odinger picture), we observe that
\[
\langle \hat\rho_0, i\hbar  \widehat{{\gamma}}\rangle= \langle\psi_0 , i\hbar\eta^{*}(  {\cal U}^\dagger\de {\cal U} )\psi_0 \rangle= \langle U\psi_0  , i\hbar\de U \psi_0 \rangle= {\mathcal{A}}_{B,0}-\eta^*{\mathcal{A}}_B,
\]
where ${\mathcal{A}}_{B,0}=\langle\psi_0,-i\hbar \de\psi_0\rangle$ and we recall ${\cal U}=\eta_* U$. Thus, if we introduce the Berry curvature ${\mathcal{B}}=\de{\mathcal{A}}_B$  and {denote} $\Omega=\eta^*\omega$, taking the exterior differential of \eqref{eq:consv1} yields
\beq
\frac{\de}{\de t}\eta^*(\omega+{\mathcal{B}})=0,
\label{PoinInv}
\eeq
which recovers the relation ${\omega{+}{\mathcal{B}}=\eta_*(\omega{+}{\mathcal{B}}_0)}$ {already}  found  in the  Schr\"odinger picture \cite{GBTr22a,GBTr22}.
\end{remark}

At this point, using the last two in \eqref{eq:equationConvective} along with Cartan's magic formula $\pounds_{{\cal X}}{A}={{\cal X}}
    \lrcorner\,
    \de{A}+\de({{\cal X}}
    \lrcorner\,
    {A})$, we obtain
\begin{equation}
    \label{eq:finaleq1}
    {{\cal X}}
    \lrcorner
    \big(\de\Theta
    +
    \langle
            \hat\rho_0
        ,
            i\hbar 
                \de^{\widehat{{\gamma}}} \widehat{{\gamma}} 
    \rangle\big)
    =
    -
    \langle
\hat\rho_0
        ,
            \de^{\widehat{{\gamma}}}\hmH
    \rangle.
\end{equation}
Here, the symbol $\lrcorner$ denotes the tensor insertion of vector fields into differential forms. Crucially, we remember that  $\widehat{{\gamma}}$ is a flat connection, so that its curvature vanishes, that is $\de^{\widehat{{\gamma}}} \widehat{{\gamma}} =0$. Thus, upon using the notation $\langle\widehat{A}\rangle=\langle\hat\rho_0,\widehat{A}\rangle$ and introducing the \emph{convective symplectic form} $\Omega=-\de\Theta$, we have
\beq
    {{\cal X}}\lrcorner\,\Omega=\langle\de^{\widehat{{\gamma}}}\widehat{\mathcal{H}}\rangle 
    \implies 
    {{\cal X}}=\Pi\langle\de^{\widehat{{\gamma}}}\widehat{\mathcal{H}}\rangle,
    \label{vectFldEhr}
\eeq
where 
\beq\label{PoissonBV}
\Pi=-\Omega^{-1}=(\de\Theta)^{-1}
\eeq
is the \emph{convective Poisson tensor}, that is a contravariant two-tensor field. Then, if $\pi=-\omega^{-1}$ is the Eulerian Poisson tensor,  we  write $X_{\widehat{H}}=\pi\de\widehat{H}$ and  \eqref{vectFldEhr} implies 
\beq\label{convectVF}
{{\cal X}}=\Pi\langle\eta^*({\cal U}^\dagger\de\widehat{H}{\cal U})\rangle=\langle\eta^*({\cal U}^\dagger X_{\widehat{H}}{\cal U})\rangle
=:\langle{\cal X}_{\widehat{\cal H}}\rangle
\eeq
so that the convective vector field ${\cal X}_{\widehat{\cal H}}=\eta^*({\cal U}^\dagger X_{\widehat{H}}{\cal U})$ evolves on the orbits of the Eulerian vector field $X_{\widehat{H}}$ under the natural action of the semidirect-product group $\operatorname{Diff}(T^*Q)\,\circledS\,{\cal F}(T^*Q,\mathscr{U}(\mathscr{H}_\text{\tiny\sf Q}))$. Observe that, since  $\eta^*({\cal U}^\dagger\de\widehat{H}{\cal U})=\de^{\widehat{{\gamma}}}\widehat{\mathcal{H}}$, we have ${\cal X}_{\widehat{\cal H}}=\Pi\de^{\widehat{{\gamma}}}\widehat{\mathcal{H}}$. Here, we notice the slight abuse of notation:  $\Pi$ is used to denote both the contravariant Poisson tensor and the sharp isomorphism $\Omega^\sharp$ between one forms and vector fields, which is induced by the non-degenerate \makebox{two-form $\Omega$}.

Finally, using the above expressions for ${\cal X}$ and $\hat\zeta$ and the zero-curvature relation $\de^{\widehat{{\gamma}}}{\widehat{\boldsymbol{\gamma}}}=0$, we are left with the following set of equations of motion
\begin{align}
\frac{\partial \Omega}{\partial t}=&\ 
%\de\langle\de^{\widehat{{\gamma}}}\hmH\rangle=
\big\langle\de^{\widehat{{\gamma}}}\hat\rho_0,\wedge\de^{\widehat{{\gamma}}}\hmH\big\rangle
\\
\frac{\partial \hmH}{\partial t}=&\ \Pi(\de^{\widehat{{\gamma}}}\widehat{\mathcal{H}},{\langle
        \de^{\widehat{{\gamma}}}\widehat{\mathcal{H}}
    \rangle})\label{HamEhrEvol}
    \\
i\hbar\frac{\partial \widehat{\boldsymbol{\gamma}}}{\partial t}=&\ \de^{\widehat{{\gamma}}}\hmH,
\label{ConnEhrEvol}
\end{align}
where $\widehat{\boldsymbol{\gamma}}|_{t=0}=0$ and we recall $\Pi=-\Omega^{-1}$ as in \eqref{PoissonBV}. The first equation above follows from $\de^{\widehat{{\gamma}}}\de^{\widehat{{\gamma}}}=0$ by using \eqref{ConnEhrEvol} in \eqref{eq:consv1}, and taking the differential. 
    Here, the purely classical case is readily recovered by dropping the hat on the Hamiltonian,  that is $\widehat{\cal H}={\cal H}\boldsymbol{1}$. In this case, $\de^{\widehat{{\gamma}}}\hmH=\de{\cal H}\boldsymbol{1}$, so that ${\partial_t {\cal H}}=0$.  Also, since  $\widehat{\boldsymbol{\gamma}}|_{t=0}=0$, the equation $i\hbar{\partial_t \widehat{\boldsymbol{\gamma}}}= \de{\cal H}\boldsymbol{1}$ simply returns an $\mathscr{U}\!(1)$-gauge 
 connection given by an exact one-form whose evolution decouples entirely. Alternatively, the purely quantum case is found by setting $\de^{\widehat{{\gamma}}}\widehat{\cal H}=0$ after observing that this condition is preserved in time by the evolution equation $(\partial_t-\pounds_{{\cal X}})\de^{\widehat{{\gamma}}}\widehat{\cal H}=[\de^{\widehat{{\gamma}}}\widehat{\cal H},\hat\zeta]$ following from the first in \eqref{mario}. We emphasize that, unlike the case of purely quantum and classical dynamics, the Heisenberg equation \eqref{HamEhrEvol} for  the hybrid Hamiltonian operator does not return a simple constant of motion. This is due to the occurrence of the expected value on the right-hand side. This fact does not mean, however, that energy is not conserved. Indeed, we notice that, upon recalling the relation  ${\cal X}_{\widehat{\cal H}}=\Pi\de^{\widehat{{\gamma}}}\widehat{\mathcal{H}}$, equation \eqref{HamEhrEvol} may be written as $\partial_t\hmH=\Omega({\cal X}_{\hmH}, {\langle
        {\cal X}_{\hmH}
    \rangle})$ thereby recovering  conservation of the \emph{specific energy} $\langle\hmH\rangle=\langle\hat\rho_0,\hmH\rangle$, that is $\partial_t\langle\hmH\rangle=\Omega({\langle
        {\cal X}_{\hmH}
    \rangle},\langle{\cal X}_{\hmH}\rangle)=0$. The conservation of the total energy $\int\!f_0\langle\hmH\rangle\,\de^2z$ follows trivially. The absence of a conservation law for the Heisenberg Hamiltonian $\widehat{\cal H}$ is essentially due to the nonlinear character of the underlying Schr\"odinger dynamics and represents a peculiarity of mixed quantum-classical dynamics. This fact suggests that, unlike purely quantum and classical observables, hybrid observables do not comprise a  Lie-algebra structure. As we will see, the same feature is also present in the Koopman  model \eqref{HybEq1}-\eqref{HybEq3}.

The present treatment can be extended to other observables than the Hamiltonian. We recall the first  in \eqref{mario} and write the original Schr\"odinger Hamiltonian $\widehat{H}$ as a matrix analytic function of a certain array of observables, that is operator-valued functions $\widehat{{O}}=(\widehat{O}_1,\widehat{O}_2,\dots)$. Then, we have
\[
\widehat{\cal H}=\widehat{H}(\widehat{{\cal O}})
\qquad\text{with}\qquad
\widehat{{\cal O}}=({\cal U}^\dagger\widehat{{O}}{\cal U})\circ\boldsymbol\eta
\]
%With these definitions, the Heisenberg Lagrangian reads
%\[
%\ell_{\sf H}(\boldsymbol{\cal X},\boldsymbol{\Theta},\hat\zeta,\widehat{\boldsymbol{\gamma}}, \widehat{\boldsymbol{\cal O}})=\!\int\!\! f_0\big(\boldsymbol{\Theta}\cdot\boldsymbol{\cal X}-\big\langle i\hbar\hat\rho_0,\hat\zeta+\boldsymbol{\cal X}\cdot\widehat{\boldsymbol{\gamma}}-\widehat{H}(\widehat{\boldsymbol{\cal O}})\big\rangle\big)\de^2z,
%\]
so that the Heisenberg dynamics of the Hamiltonian is replaced by the corresponding dynamics of the observables $\widehat{{\cal O}}=(\widehat{\cal O}_1,\widehat{\cal O}_2,\dots)$ obeying the equation $\partial_t\widehat{{\cal O}}+[\hat\zeta,\widehat{{\cal O}}]=\pounds_{\cal X}\widehat{{\cal O}}$. Making use of the relations \eqref{infgen} and \eqref{vectFldEhr}, the Heisenberg dynamics of quantum-classical observables becomes
\begin{equation}
\label{eq:observableehr}
\frac{\partial \widehat{{\cal O}}}{\partial t}=\Pi(\de^{\widehat{{\gamma}}} \widehat{{\cal O}},{\langle
        \de^{\widehat{{\gamma}}}\widehat{\mathcal{H}}
    \rangle})+i\hbar^{-1}[\hmH,\widehat{{\cal O}}],
\end{equation}
so that the local expectation value $\langle\widehat{{\cal O}}\rangle=\langle\hat\rho_0,\widehat{{\cal O}}\rangle$ evolves according to
\[
\frac{\partial \langle\widehat{{\cal O}}\rangle}{\partial t}=\Pi(\langle\de^{\widehat{{\gamma}}} \widehat{{\cal O}}\rangle,{\langle
        \de^{\widehat{{\gamma}}}\widehat{\mathcal{H}}
    \rangle})+\hbar^{-1}\langle i[\hmH,\widehat{{\cal O}}]\rangle.
\]
In the purely classical case, we have $\widehat{\cal H}={\cal H}\boldsymbol{1}$ and $\widehat{\boldsymbol{\gamma}}={\boldsymbol{\gamma}}\boldsymbol{1}$, so that $\partial_t\Omega=0$ and ${\cal H}=H$. We recognize the usual  evolution ${\partial_t {{\cal O}}}=\omega^{-1}(\de {{\cal O}},{
        \de{{H}}
    })=\{{{\cal O}},{H}\}$ for the classical observable $\widehat{{\cal O}}={{\cal O}}\boldsymbol{1}$. Once again, the purely quantum case is recovered by setting $\de^{\widehat{{\gamma}}}\widehat{\mathcal{H}}=0$. {We emphasize that the local expectation values $\langle\widehat{O}\rangle$ must not be confused with the overall expectation values ${\int} f_0\langle\widehat{O}\rangle\de^2 z_0$, which are the measurable quantities. As we will see later, the overall expectation may be conserved even if the local expectation is not.}

\begin{remark}[Special cases and conservation laws]\label{rem:ConsLaws}
Another way to describe \eqref{eq:observableehr} is by rewriting the latter as
\begin{equation}
    \frac{\partial \widehat{{\cal O}}}{\partial t}=\Omega(\mathcal{X}_{\widehat{\mathcal{O}}},{{\langle
        \mathcal{X}_{\widehat{\mathcal{H}}}
    \rangle}})+i\hbar^{-1}[\hmH,\widehat{{\cal O}}].
    \end{equation}
Here we can distinguish three special cases.
 In the first case, $\Omega(\mathcal{X}_{\widehat{\mathcal{O}}},{{\langle
        \mathcal{X}_{\widehat{\mathcal{H}}}
    \rangle}})=0$, one says that $\mathcal{X}_{\widehat{\mathcal{O}}}$ and $\langle
        \mathcal{X}_{\widehat{\mathcal{H}}}
    \rangle$ are \emph{symplectically orthogonal} and the equation of motion reduces to the Heisenberg equation ${\partial_t \widehat{{\cal O}}}=i\hbar^{-1}[\hmH,\widehat{{\cal O}}]$. The symplectic orthogonality condition  is written in the Eulerian frame as $0=\omega({X}_{\widehat{{O}}},{{\langle
        {X}_{\widehat{{H}}}
    \rangle}})=\pounds_{\langle X_{\widehat{{H}}}
    \rangle}{\widehat{{O}}}$. Such an observable is unaffected by the action of the diffeomorphism in the evolution. This is the case for a purely quantum observable $\widehat{\cal O}=\eta^*({\cal U}^\dagger\widehat{O}{\cal U})$. 
    %If instead this observable is purely classical, that is $\widehat{{\cal O}}=\mathcal{O}=\eta_*{O}$, then it appears as a constant of motion.
 In the second case, we have $[\hmH,\widehat{{\cal O}}]=0$, so that
        ${\partial_t \widehat{{\cal O}}}=\Omega(\mathcal{X}_{\widehat{\mathcal{O}}},{{\langle
        \mathcal{X}_{\widehat{\mathcal{H}}}
    \rangle}})$.
        In this case the apparent quantum evolution of the observable is determined by the presence of the connection $\widehat\gamma$ in the construction of the vector fields. The appearance of $\widehat\gamma$, however, is simply related with the gauge nature of the quantum frame induced by $\mathcal{U}$  and should not be understood as a proper quantum evolution. This is the case for a purely classical observable. 
%If instead, this  observable is purely quantum, one has $\de^{\widehat\gamma}\widehat{\mathcal{O}}=0$, so that $\widehat{\mathcal{O}}$ appears as a constant of motion. 
In the third case, both $\Omega(\mathcal{X}_{\widehat{\mathcal{O}}},{{\langle
        \mathcal{X}_{\widehat{\mathcal{H}}}
    \rangle}})=0$ and $[\hmH,\widehat{{\cal O}}]=0$, so that the  operator $\widehat{{\cal O}}$ is conserved.  
    %As we saw before, a     {more common occurrence} is that both ${\Omega(\langle\mathcal{X}_{\widehat{\mathcal{O}}}\rangle,{{\langle\mathcal{X}_{\widehat{\mathcal{H}}}\rangle}})=0}$ and $[\hmH,\widehat{{\cal O}}]=0$ hold, in which case the local expectation $\langle{\widehat{\mathcal{O}}}\rangle$ is conserved.

    {One can also consider the conservation of local expectations, which may result from the combination of  different properties. For example, the local expectation $\langle\widehat{\cal N}\rangle$ of the quantity $\widehat{\cal N}=({\mathcal{P}}^2+{\mathcal{Q}}^2+\widehat{{\Sigma}}_z)/2$ is conserved by the Jaynes-Cummings Hamiltonian $\widehat{\mathcal{H}}=\widehat{\cal N}+g({\mathcal{Q}}\widehat{{\Sigma}}_x-{\mathcal{P}}\widehat{{\Sigma}}_y)$, where  $({\cal Q},{\cal P})=(\eta^*q,\eta^*p)$ and, in the usual Pauli matrix notation, $\widehat{\boldsymbol\Sigma}=\eta^*({\cal U}^\dagger\widehat{\boldsymbol\sigma}{\cal U})$. This conservation is verified directly upon using $0=\eta^*({\cal U}^\dagger\de\widehat{\boldsymbol\sigma}{\cal U})=\de^{\widehat{\gamma}}\widehat{\boldsymbol\Sigma}$ so that, for example, $\mathcal{X}_{{\mathcal{Q}}\widehat{{\Sigma}}_x}=\widehat{{\Sigma}}_x\mathcal{X}_{{\mathcal{Q}}}$. Similarly, one shows the conservation of the expectation $\langle\widehat{\cal M}\rangle$ of the total momentum $\widehat{\cal M}={\mathcal{P}}{+}\,\widehat{\!\mathscr{K}}$ for translation invariant Hamiltonians of the type $\widehat{\mathcal{H}}=(M^{-1}{\mathcal{P}}^2+m^{-1}\widehat{\!\mathscr{K}\,}\!^2)/2+V(\mathcal{Q}-\widehat{\!\mathscr{X}})$, where $(\widehat{\!\mathscr{X}},\widehat{\!\mathscr{K}})=\eta^*({\cal U}^\dagger(\widehat{x},\widehat{p}){\cal U})$ are quantum canonical observables such that $[\widehat{\!\mathscr{X}},\widehat{\!\mathscr{K}}]=i\hbar$. Here, the conservation of $\langle\widehat{\cal M}\rangle$ follows from $\de^{\widehat{{\gamma}}\,}\widehat{\!\mathscr{X}}=0$, so that $\de^{\widehat{{\gamma}}}V=\partial_{\mathcal{Q}}V \de\mathcal{Q}$ and $\Omega(\mathcal{X}_{\widehat{\mathcal{O}}},{{\langle
        \mathcal{X}_{\widehat{\mathcal{H}}}
    \rangle}})=\Pi(\de {{\cal P}},{\langle
        \de^{\widehat{{\gamma}}}\widehat{\mathcal{H}}
    \rangle})=\hbar^{-1}\langle i[\,\widehat{\!\mathscr{K}},\hmH]\rangle$.
\noindent
}

\end{remark}

\section{Heisenberg picture of hybrid Koopman dynamics}
\label{sec:Koopman}

In this section, we extend the previous treatment to the Koopman model  \eqref{HybEq1}-\eqref{HybEq3} by including the backreaction term occurring in the Hamiltonian functional \eqref{KoopHam}. By proceeding analogously to the beginning of Section \ref{sec:Ehrenfest}, we write the Lagrangian given by \eqref{MQCLagr} and \eqref{KoopHam} as follows:
\[
\ell_{\sf H}(\boldsymbol{\cal X},\boldsymbol{\Theta},\hat\zeta,\widehat{\boldsymbol{\gamma}}, \widehat{\cal H})=\!\int\!\! f_0\left(\boldsymbol{\Theta}\cdot\boldsymbol{\cal X}+\big\langle \hat\rho_0,i\hbar(\hat\zeta+\boldsymbol{\cal X}\cdot\widehat{\boldsymbol{\gamma}})-\widehat{\cal H}\big\rangle-\frac12
    \big\langle
        \bbX_{\hmH}
        ,
            {i\hbar}
            \big[
                    \hat\rho_0
                ,
                    \nabla^{\widehat{{\gamma}}}\hat\rho_0
                                \big]
    \big\rangle\right) 
    \de^2z_0,
\]
where we recall ${\cal X}_{\widehat{\cal H}}:=(\de\Theta)^{-1}\de^{\widehat{{\gamma}}}\widehat{\mathcal{H}}$. From \eqref{convectVF}, we also recall that the latter quantity evolves on orbits of $X_{\widehat{H}}$ under the  action of the semidirect-product group $\operatorname{Diff}(T^*Q)\,\circledS\,{\cal F}(T^*Q,\mathscr{U}(\mathscr{H}_\text{\tiny\sf Q}))$, that is $X_{\widehat{H}}\mapsto\eta^*({\cal U}^\dagger X_{\widehat{H}}{\cal U})={\cal X}_{\widehat{\cal H}}$. Thus, for the sake of writing the dynamics, it is convenient to add the latter quantity, now denoted by $\widehat{\cal Z}=\bZ\cdot\nabla$, to the set of dynamical variables, so that the Lagrangian becomes
\[
\ell_{\sf H}(\boldsymbol{\cal X},\boldsymbol{\Theta},\hat\zeta,\widehat{\boldsymbol{\gamma}}, \widehat{\cal H},\bZ)=\!\int\!\! f_0\left(\boldsymbol{\Theta}\cdot\boldsymbol{\cal X}+\big\langle \hat\rho_0,i\hbar(\hat\zeta+\boldsymbol{\cal X}\cdot\widehat{\boldsymbol{\gamma}})-\widehat{\cal H}\big\rangle-\frac12
    \big\langle
        \bZ
        ,
            {i\hbar}
            \big[
                    \hat\rho_0
                ,
                    \nabla^{\widehat{{\gamma}}}\hat\rho_0
                                \big]
    \big\rangle\right) 
    \de^2z_0,
\]
with
\[
    \widehat{\cal Z} = \eta^*({\cal U}^\dagger X_{\widehat{H}}{\cal U}),
    \qquad \ 
    \partial_t \widehat{\cal Z}=\pounds_{{\cal X}}
\widehat{\cal Z}
+
[\widehat{\cal Z},\hat\zeta]
,\qquad \ 
\delta\widehat{\cal Z}=\pounds_{{\cal Y}}
\widehat{\cal Z}
+
[\widehat{\cal Z},\hat\nu].
\]
Here, the last two follow by taking the relevant derivative of the first. 
Notice that, since $\operatorname{div} X_{\widehat{H}}=0$, we have
\beq\label{zerodiv}
0=\eta^*({\cal U}^\dagger(\operatorname{div} X_{\widehat{H}}){\cal U})
=
\eta^*\operatorname{div}({\cal U}^\dagger X_{\widehat{H}}{\cal U})
+
\eta^*[{\cal U}^\dagger\de{\cal U},{\cal U}^\dagger X_{\widehat{H}}{\cal U}]
=
\operatorname{div}\widehat{\cal Z}+[\widehat{\boldsymbol{\gamma}},\widehat{\cal Z}]=:\operatorname{div}^{\widehat{\gamma}}\widehat{\cal Z},
\eeq
where the last step defines the usual covariant divergence operator.

With the inclusion of this quantity in the set of dynamical variables, equation \eqref{eq:quantehr} becomes
\beq
\frac{\partial}{\partial t}\frac{\delta \ell_{\sf H}}{\delta \hat\zeta}+\left[\hat\zeta,\frac{\delta \ell_{\sf H}}{\delta \hat\zeta}\right]=\operatorname{div}\!\left(\frac{\delta \ell_{\sf H}}{\delta \hat\zeta}\boldsymbol{\cal X}\right) 
- \operatorname{div}^{\widehat{{\gamma}}}\!\left(\frac{\delta \ell_{\sf H}}{\delta \widehat{\boldsymbol{\gamma}}}\right)
%-
%\left[\widehat{\boldsymbol{\gamma}},\frac{\delta \ell_{\sf H}}{\delta \widehat{\boldsymbol{\gamma}}}\right]
+
\left[\hmH,\frac{\delta \ell_{\sf H}}{\delta \hmH}\right]
+
\left[\bZ,\frac{\delta \ell_{\sf H}}{\delta \bZ}\right],
\label{eq:quantehr2}
\eeq
where we evaluate
\beq
        \frac{\delta \ell_{\sf H}}{\delta
        \hat\zeta}= -i\hbar f_0\hat\rho_0
                \,,\qquad
        \frac{\delta \ell_{\sf H}}{\delta
        \hmH}= -f_0\hat\rho_0
,
\label{funders1}
    \eeq
and
    \beq
        \frac{\delta \ell_{\sf H}}{\delta
        \widehat{\boldsymbol{\gamma}}}= -i\hbar f_0\Big(\hat\rho_0\boldsymbol{\cal X}
        +        \frac{1}{2}
        \big[
            \hat\rho_0
        ,
        [
            \hat\rho_0
            ,
            \bZ
        ]
    \big]\Big)
        \,,\qquad    
        \frac{\delta \ell_{\sf H}}{\delta \bZ}=-\frac{i\hbar}{2}
        f_0
        \big[
                    \hat\rho_0
                ,
                    \nabla^{\widehat{{\gamma}}}\hat\rho_0
                                \big]
    .
    \label{funders2}
\eeq
In particular, using the Leibniz property of the covariant divergence as well as the Jacobi identity, we compute
\begin{align*}
    \operatorname{div}^{\widehat{{\gamma}}}\!
    \left(
        f_0
        \Big[
            \hat\rho_0
            ,
            \big[
                \hat\rho_0
                ,
                \bZ
            \big]
        \Big]
    \right)
    -f_0
        \Big[
        \bZ
        ,
            \big[
                    \hat\rho_0
                ,
                    \nabla^{\widehat{{\gamma}}}\hat\rho_0
            \big]
    \Big]
   = &\ 
        f_0
        \Big[
            \nabla^{\widehat{{\gamma}}}\hat\rho_0
            ,
            \big[
                \hat\rho_0
                ,
                \bZ
            \big]
        \Big]
        +
        f_0
        \Big[
            \hat\rho_0
            ,
            \big[
                \nabla^{\widehat{{\gamma}}}\hat\rho_0
                ,
                \bZ
            \big]
        \Big]
                \\
 &
        +\nabla f_0\cdot
        \Big[
            \hat\rho_0
            ,
            \big[
                \hat\rho_0
                ,
                \bZ
            \big]
        \Big]
              -
              f_0
            \Big[
        \bZ
        ,
            \big[
                    \hat\rho_0
                ,
                    \nabla^{\widehat{{\gamma}}}\hat\rho_0
            \big]
    \Big]
        \\
=        &\ 
        f_0
        \Big[
            \hat\rho_0
            ,
            \big[2
                \nabla^{\widehat{{\gamma}}}\hat\rho_0
                +
                \hat\rho_0{\nabla {\ln} f_0}
                ,
                \bZ
            \big]
        \Big],
\end{align*}
so that the equation for the quantum generator $\hat\zeta$ reads
\[
\Big[\hat\rho_0,i\hbar(\hat\zeta+\boldsymbol{\cal X}\cdot\widehat{\boldsymbol{\gamma}})-\hmH-           i\hbar \big[
                \nabla^{\widehat{{\gamma}}}\hat\rho_0
                +
                \hat\rho_0{\nabla {\ln} \sqrt{f_0}}
                ,
                \bZ
            \big]
        \Big]=0,
\]
where the notation is such that $[
                \nabla^{\widehat{{\gamma}}}\hat\rho_0
                +
                \hat\rho_0{\nabla {\ln} \sqrt{f_0}},\bZ]=[
                \partial_k^{\widehat{{\gamma}}}\hat\rho_0
                +
                \hat\rho_0{\partial_k {\ln} \sqrt{f_0}},\widehat{\cal Z}^k]$. 
Upon recalling $\bZ={\bbX}_{\widehat{\cal H}}$, this implies
\begin{align}
\hat\zeta=&\ -\boldsymbol{\cal X}\cdot\widehat{\boldsymbol{\gamma}}-i\hbar^{-1}\hmH
+\big[
                \nabla^{\widehat{{\gamma}}}\hat\rho_0
                ,
                \bbX_{\widehat{\cal H}}
            \big]
+\big[
                \hat\rho_0
                ,
                 \bbX_{\widehat{\cal H}}
            \big]\cdot\nabla {\ln}\sqrt{f_0}+\hat\zeta_0
%            \nonumber\\
%=&\ -\boldsymbol{\cal X}\cdot\widehat{\boldsymbol{\gamma}}-i\hbar^{-1}\hmH
%+\frac1{\sqrt{f_0}}\big[
%                \de^{\widehat{{\gamma}}}(\sqrt{f_0}\hat\rho_0)
%                ,
%                \bbX_{\widehat{\cal H}}
%            \big]+\hat\zeta_0
            \label{infgen2}
,
\end{align}
where 
$[\hat\rho_0,\hat\zeta_0]=0$ and again we  set the gauge $\hat\zeta_0=0$.

\subsection{The convective vector field}
\label{sec:KoopmanVectorField}

We now move on to consider the equations produced by the  variations ${\boldsymbol{\cal Y}=\eta^*(\delta\boldsymbol\eta\circ\boldsymbol\eta^{-1})}$, that is
\begin{align}\nonumber
\frac{\partial}{\partial t}\frac{\delta \ell_{\sf H}}{\delta
\boldsymbol{\cal X}}
=&\, 
\pounds_{\boldsymbol{\cal X}}\,\frac{\delta \ell_{\sf H}}{\delta \boldsymbol{\cal X}}
+
\Big\langle\frac{\delta \ell_{\sf H}}{\delta
{\hmH}},\nabla{\hmH}\Big\rangle+\Big\langle\frac{\delta \ell_{\sf H}}{\delta
\hat\zeta},\nabla{\hat\zeta}\Big\rangle
+
\nabla\boldsymbol\Theta\cdot\frac{\delta \ell_{\sf H}}{\delta
\boldsymbol\Theta}
-
\boldsymbol\Theta\operatorname{div}\frac{\delta \ell_{\sf H}}{\delta
\boldsymbol\Theta}
-\frac{\delta \ell_{\sf H}}{\delta
\boldsymbol\Theta}\cdot\nabla\boldsymbol\Theta 
\nonumber
\\
&
+\Big\langle\nabla\widehat{\boldsymbol{\gamma}},\frac{\delta \ell_{\sf H}}{\delta
\widehat{\boldsymbol{\gamma}}}\Big\rangle
-
\Big\langle\widehat{\boldsymbol{\gamma}},\operatorname{div}\frac{\delta \ell_{\sf H}}{\delta
\widehat{\boldsymbol{\gamma}}}\Big\rangle
-
\Big\langle\frac{\delta \ell_{\sf H}}{\delta
\widehat{\boldsymbol{\gamma}}}\cdot\nabla,\widehat{\boldsymbol{\gamma}}\Big\rangle
-
\Big\langle\pounds_{\bZ},\frac{\delta \ell_{\sf H}}{\delta
    \bZ}\Big\rangle,
%    \nonumber
%\\
%&
%-
%    \Big\langle\nabla\bZ,\frac{\delta \ell_{\sf H}}{\delta
%    \bZ}\Big\rangle
%    -
%        \Big\langle\frac{\delta \ell_{\sf H}}{\delta
%    \bZ},\operatorname{div}\bZ\Big\rangle
%    -
%    \Big\langle\bZ\cdot\nabla,\frac{\delta \ell_{\sf H}}{\delta
%\bZ}\Big\rangle
\label{eq:ehrconv2}
\end{align}
where we have introduced the convenient notation
\[
\langle\pounds_{\widehat{\boldsymbol{\cal W}}},\widehat{\boldsymbol\alpha}\rangle
:=
%\langle\widehat{\boldsymbol{\cal W}}\cdot\nabla,\widehat{\boldsymbol\alpha}\rangle
\partial_k\langle\widehat{{\cal W}}^k,\widehat{\alpha}\rangle
+
\langle\widehat{\alpha}_k,\nabla\widehat{{\cal W}}^k\rangle
,
\]
for any operator-valued vector field $\widehat{{\cal W}}=\widehat{\boldsymbol{\cal W}}\cdot\nabla$ and any operator-valued one-form density $\widehat{\alpha}=\widehat{\boldsymbol\alpha}\cdot\de\bz\otimes\de^2z$.
Upon  evaluating ${\delta \ell_{\sf H}}/{\delta
        \boldsymbol{\cal X}}= f_0\boldsymbol{\boldsymbol\Theta} + f_0\langle \hat\rho_0,i\hbar\widehat{\boldsymbol{\gamma}} \rangle$, ${\delta \ell_{\sf H}}/{\delta \boldsymbol\Theta}= f_0\boldsymbol{\cal X}$, and recalling  \eqref{funders1}-\eqref{funders2}, we obtain
\begin{align}
\nonumber
\frac{\partial}{\partial t}\big({\boldsymbol\Theta}+\langle \hat\rho_0,i\hbar \widehat{\boldsymbol{\gamma}} \rangle\big)
=&\, 
\nabla\Big(\boldsymbol{\cal X}\cdot{\boldsymbol\Theta} +
\frac12
            \big\langle
                {                    \bZ
            ,
            {i\hbar}
            [
                    \rho_0
                ,
                [
                    \rho_0
                    ,
            \widehat{\boldsymbol{\gamma}}
                            ]
            ]}
            \big\rangle  \Big)
+\frac12
\big\langle \hat\rho_0,i\hbar\nabla \big[
                2\nabla^{\widehat{{\gamma}}}\hat\rho_0
                +
                \hat\rho_0{\nabla {\ln} {f_0}}
                ,
                \bZ
            \big]
\big\rangle
\\
&-\frac1{2f_0}
            \Big\langle
                \pounds_{
                    [
                    \hat\rho_0
                ,
                [
                    \hat\rho_0
                    ,
                    \bZ
                ]
            ]}
            ,
            {i\hbar}f_0
            \widehat{\boldsymbol{\gamma}}
            \Big\rangle
            -\frac1{2f_0}
        \Big\langle
            \pounds_{\bZ}
            ,
            {i\hbar}f_0
                [
                        \hat\rho_0
                    ,
                    \nabla^{\widehat{\boldsymbol{\gamma}}}\hat\rho_0
%                        [
%                            \hat\rho_0
%                            ,
%                            \widehat{\boldsymbol{\gamma}}
%                        ]
                ]
        \Big\rangle,
\label{eq:consv2}
\end{align}
Here, we have used 
%$
%\nabla\boldsymbol{\Theta}\cdot{\delta \ell_{\sf H}}/{\delta
%\boldsymbol{\Theta}}
%-
%\partial_k({\delta \ell_{\sf H}}/{\delta
%{\Theta}}_k\boldsymbol{\Theta})
%=
%f_0\nabla(f_0^{-1}\boldsymbol{\Theta}\cdot {\delta \ell_{\sf H}}/{\delta
%\boldsymbol{\Theta}})-\pounds_{f_0^{-1}{\delta \ell_{\sf H}}/{\delta
%\boldsymbol{\Theta}}}(f_0\boldsymbol{\Theta})
%$
%and, analogously,
$
\big\langle\nabla\widehat{{\gamma}}_k,{\delta \ell_{\sf H}}/{\delta
\widehat{{\gamma}}}_k\big\rangle
-
\partial_k\big\langle{\delta \ell_{\sf H}}/{\delta
\widehat{{\gamma}}}_k,\widehat{\boldsymbol{\gamma}}\big\rangle
%\\
%=
%f_0\nabla\big\langle\widehat{\boldsymbol{\gamma}},f_0^{-1}{\delta \ell_{\sf H}}/{\delta
%\widehat{\boldsymbol{\gamma}}}\big\rangle
%-
%\big\langle\nabla (f_0^{-1}{\delta \ell_{\sf H}}/{\delta
%\widehat{\boldsymbol{\gamma}}}),f_0\widehat{\boldsymbol{\gamma}}\big\rangle
%-
%\operatorname{div}\big\langle{\delta \ell_{\sf H}}/{\delta
%\widehat{\boldsymbol{\gamma}}},\widehat{\boldsymbol{\gamma}}\big\rangle
%\\
=
f_0\nabla\big\langle f_0^{-1}{\delta \ell_{\sf H}}/{\delta
\widehat{{\gamma}}}_k,\widehat{{\gamma}}_k\big\rangle-\big\langle\pounds_{f_0^{-1}{\delta \ell_{\sf H}}/{\delta
\widehat{\boldsymbol{\gamma}}}},f_0\widehat{\boldsymbol{\gamma}}\big\rangle
$ and an analogous relation for the last three terms in the first line of \eqref{eq:ehrconv2}.

By letting the operation $[\bZ,\widehat{\boldsymbol{\alpha}}]=[\widehat{\cal Z}^k,\widehat{\alpha}_k]$ have priority, we calculate
\begin{align*}
\Big\langle
                \pounds_{
                    [
                    \hat\rho_0
                ,
                [
                    \hat\rho_0
                    ,
                    \bZ
                ]
            ]}
            ,
            {i }f_0
            \widehat{\boldsymbol{\gamma}}
            \Big\rangle
            \!+\!
        \Big\langle
            \pounds_{\bZ}
            ,
            {i }f_0
                [
                        \hat\rho_0
                    ,
                        \nabla^{\widehat{\boldsymbol{\gamma}}}
                            \hat\rho_0
                ]
        \Big\rangle
        =&\, f_0
           \big\langle{i }
            \widehat{{\gamma}}_k,
                \nabla{
                    [
                    \hat\rho_0
                ,
                [
                    \hat\rho_0
                    ,
                    {\cal Z}^k
                ]
            ]}
            \big\rangle
            -f_0
        \big\langle
            {i }
                [
                        \hat\rho_0
                    ,
                        [
                            \hat\rho_0
                            ,
                            \widehat{{\gamma}}_k
                        ]
                ]
                ,
            \nabla{\cal Z}^k
        \big\rangle    
\\        
        &+
\partial_k       \Big(    \big\langle
                {
                    [
                    \hat\rho_0
                ,
                [
                    \hat\rho_0
                    ,
                    {\cal Z}^k
                ]
            ]}
            ,
            {i }f_0
            \widehat{\boldsymbol{\gamma}}
            \big\rangle
            - 
    \big\langle
            {\cal Z}^k
            ,
            {i }f_0
                [
                        \hat\rho_0
                    ,
                        [
                            \hat\rho_0
                            ,
                                                        \widehat{\boldsymbol{\gamma}}
                        ]
                ]
        \big\rangle    \Big)
\\        
        &+
            f_0
        \big\langle
            {i }
                [
                        \hat\rho_0
                    ,
                        \partial_k
                            \hat\rho_0
                ]
                ,
            \nabla{\cal Z}^k
        \big\rangle    
        +
                    \partial_k
        \big\langle
            {\cal Z}^k
            ,
            {i }f_0
                [
                        \hat\rho_0
                    ,
                        \nabla
                            \hat\rho_0
                ]
        \big\rangle    
\\        =&\, 
           f_0\big\langle
                       {i }
            \widehat{{\gamma}}_k
            ,
                {
                    [
                    \nabla\hat\rho_0
                ,
                [
                    \hat\rho_0
                    ,
                    {\cal Z}^k
                ]
            ]}
            \big\rangle
+
           f_0\big\langle
                       {i }
            \widehat{{\gamma}}_k
            ,
                {
                    [
                    \hat\rho_0
                ,
                [
                    \nabla\hat\rho_0
                    ,
                    {\cal Z}^k
                ]
            ]}
            \big\rangle
\\        
        &+
            f_0\nabla
        \big\langle
            {\cal Z}^k
            ,
            {i }
                [
                        \hat\rho_0
                    ,
                        \partial_k
                            \hat\rho_0
                ]
        \big\rangle    
-            f_0
        \big\langle
                           {\cal Z}^k
                            ,
            {i }
                [\partial_k
                        \hat\rho_0
                    ,
                        \nabla
                            \hat\rho_0
                ]
        \big\rangle    
\\
        &
        +
        \big\langle
\operatorname{div}            {\bZ}
            ,
            {i }f_0
                [
                        \hat\rho_0
                    ,
                        \nabla
                            \hat\rho_0
                ]
        \big\rangle    
        +
        \big\langle                
                {\cal Z}^k
                ,
            {i }
                [                                                    \partial_k(f_0
                        \hat\rho_0)
                ,
                \nabla
                            \hat\rho_0
                                            ]
        \big\rangle    
\\=&\,
        f_0
        \big\langle
            \nabla\rho_0
                \,
            ,
            {i }
            \big(
                [
                \bZ
                ,
                [
                    \widehat{\boldsymbol{\gamma}}
                    ,
                    \hat\rho_0
                ]
                ]
            -
            [
                               [
                    \bZ
                    ,
                    \hat\rho_0
                ]
                ,
                 \widehat{\boldsymbol{\gamma}}
                ]
            \big)
        \big\rangle
\\
&+
            f_0\nabla
        \big\langle
            {\bZ}
            ,
            {i }
                [
                        \hat\rho_0
                    ,
                        \nabla
                            \hat\rho_0
                ]
        \big\rangle    
+
        \big\langle
\operatorname{div}            {\bZ}
            ,
            {i }f_0
                [
                        \hat\rho_0
                    ,
                        \nabla
                            \hat\rho_0
                ]
        \big\rangle    
\\
        &
        +
        \big\langle                
                {\bZ}\cdot\nabla f_0
                ,
            {i }
                [                                                    
                        \hat\rho_0
                ,
                \nabla
                            \hat\rho_0
                            ]
        \big\rangle    
\\
=&\,
\big\langle
    \nabla\hat\rho_0
        \,
    ,
    {i }
        [
        \bZ
        ,
        2f_0\nabla^{\widehat{{\gamma}}}\hat\rho_0
        +
        \hat\rho_0
        {\nabla f_0}
        ]
        -
        {i }
        {[
            \operatorname{div}^{\widehat{{\gamma}}}\bZ
            ,
            \hat\rho_0
            ]}
\big\rangle
\\
&
-
f_0
\nabla
\big\langle\hat\rho_0,
    {i }
    {[
        \bZ
    ,
\nabla
        \hat\rho_0
]}
\big\rangle
\\
=&\,
\big\langle
    \nabla\hat\rho_0
        \,
    ,
    {i }
        [
        \bZ
        ,
        2f_0\nabla^{\widehat{{\gamma}}}\hat\rho_0
        +
        \hat\rho_0
                {\nabla f_0}
        ]
\big\rangle
\!-\!
f_0
\nabla
\big\langle\hat\rho_0,
    {i }
    {[
        \bZ
    ,
\nabla
        \hat\rho_0
]}
\big\rangle
,
\end{align*}
where the last equality follows from \eqref{zerodiv}. Thus, by replacing $\bZ={\bbX}_{\widehat{\cal H}}$, \eqref{eq:consv2} becomes, in coordinates,
\begin{align}
\nonumber
\frac{\partial}{\partial t}\big(\boldsymbol{\Theta}+\langle \hat\rho_0,i\hbar \widehat{\boldsymbol{\gamma}} \rangle\big)
=&\, 
\nabla\Big(\boldsymbol{\cal X}\cdot\boldsymbol{\Theta} -
\frac12
\big\langle                \bbX_{\widehat{\cal H}}
,
    {i \hbar}
    {[
        \hat\rho_0
    ,
\nabla^{\widehat{{\gamma}}}
        \hat\rho_0
]}
            \big\rangle  \Big)
.
%\label{eq:consv3}
\end{align}
Taking the differential on both sides, we observe that the relation \eqref{PoinInv} applies to both Ehrenfest and Koopman dynamics \cite{GBTr22a,GBTr22}. At this point, upon denoting $\de\boldsymbol\Theta=\nabla\boldsymbol\Theta-(\nabla\boldsymbol\Theta)^T$ and ${\boldsymbol{\cal X}}
    \lrcorner
    \de\boldsymbol\Theta=\boldsymbol{\cal X}\cdot\nabla\boldsymbol\Theta-\nabla\boldsymbol\Theta\cdot\boldsymbol{\cal X}$, one makes use of   the last two in \eqref{eq:equationConvective} to obtain
\begin{align*}
    -{\boldsymbol{\cal X}}
    \lrcorner\,
    \de\boldsymbol\Theta
=&\,
  \Big  \langle
\hat\rho_0
        ,
            \nabla^{\widehat{{\gamma}}}\Big(\hmH-i\hbar \Big[
            \bZ,
                \nabla^{\widehat{{\gamma}}}\hat\rho_0
                +\frac12
                \hat\rho_0{\nabla {\ln} {f_0}}
            \Big]\Big)
   \Big \rangle
+\frac12
   \nabla\langle
\hat\rho_0,
i\hbar[\bZ,\nabla^{\widehat{{\gamma}}}\hat\rho_0]
\rangle
\\
    =&\,
      \langle
\hat\rho_0
        ,
            \nabla^{\widehat{{\gamma}}}\hmH \rangle
            -\frac12\nabla
    \langle
   \hat\rho_0
        ,            i\hbar[\bZ,
                \nabla^{\widehat{{\gamma}}}\hat\rho_0
]
             \rangle
   +
     \Big  \langle
\nabla^{\widehat{{\gamma}}}\hat\rho_0
        ,
          i\hbar   \Big[                
                \bZ,
            \nabla^{\widehat{{\gamma}}}\hat\rho_0
                +\frac12
                \hat\rho_0{\nabla {\ln} {f_0}}
            \Big]
   \Big \rangle
\\
    =&\,
      \langle
\hat\rho_0
        ,
            \nabla^{\widehat{{\gamma}}}\hmH \rangle
            -\frac12\nabla
    \langle
   \hat\rho_0
        ,            i\hbar[
                \bZ,
            \nabla^{\widehat{{\gamma}}}\hat\rho_0
               ]
             \rangle
   +
       \langle\nabla^{\widehat{{\gamma}}}\hat\rho_0,
               i\hbar   [
                  \bZ,
                \nabla^{\widehat{{\gamma}}}\hat\rho_0
                 ]
    \rangle
      -\frac1{2}
       \langle \nabla^{\widehat{{\gamma}}}\hat\rho_0,
               i\hbar   [
                \hat\rho_0,
                \bZ]\cdot{  \nabla {\ln}f_0        }
    \rangle
%    \\
%    =&\,
%      \langle
%\hat\rho_0
%        ,
%            \nabla^{\widehat{{\gamma}}}\hmH \rangle
%   +\frac12
%    \langle
%        \hat\rho_0
%        ,
%    i\hbar \nabla^{\widehat{{\gamma}}}[   
%                \nabla^{\widehat{{\gamma}}}
%                {\hat\rho_0}
%                \cdot
%                ,
%                \boldsymbol{\cal X}_\hmH
%                ] 
%            \rangle
%                -\frac1{2{f_0}}
%                \langle
%              \nabla^{\widehat{{\gamma}}} \hat\rho_0
%                    ,            i\hbar[
%                            \nabla^{\widehat{{\gamma}}}
%                            (f_0\hat\rho_0)
%                            ,\cdot
%                           \bZ ]
%                         \rangle
    \\
    =&\,
      \langle
\hat\rho_0
        ,
            \nabla^{\widehat{{\gamma}}}\hmH \rangle
   +\frac12
    \langle
        \hat\rho_0
        ,
    i\hbar [   
                \nabla^{\widehat{{\gamma}}}\partial_k^{\widehat{{\gamma}}}
                {\hat\rho_0}
                ,
                \widehat{\cal Z}^k
                ] 
            \rangle
   +\frac12
    \langle
        \hat\rho_0
        ,
    i\hbar [   
                \partial_k^{\widehat{{\gamma}}}
                {\hat\rho_0}
                ,
                \nabla^{\widehat{{\gamma}}}\widehat{\cal Z}^k
                ] 
            \rangle
                -\frac1{2}
                \langle
              \nabla^{\widehat{{\gamma}}} \hat\rho_0
                    ,            i\hbar[
                            \partial_k^{\widehat{{\gamma}}}
                            \hat\rho_0
                            ,
                           \widehat{\cal Z}^k ]
                         \rangle
    \\
    &
                                         -\frac1{2}
                \langle
              \nabla^{\widehat{{\gamma}}} \hat\rho_0
                    ,            i\hbar[\hat\rho_0
                            ,
                           \pounds_{\widehat{\cal Z}}\ln f_0 ]
                         \rangle
     \\
=&\,
    \langle
        \hat\rho_0
                ,
                    \nabla^{\widehat{{\gamma}}}\hmH \rangle
                    +\frac{\hbar}{2}
            \big(
            \langle               \pounds_{\bZ} , \widehat{\boldsymbol\Gamma}
                    \rangle
	   +    \langle \widehat{\cal Z}^k, [{\widehat{\gamma}}_k, \widehat{\boldsymbol\Gamma}]\rangle
	   -    \langle [ \widehat{\cal Z}^k, \widehat{\Gamma}_k],{\widehat{\boldsymbol\gamma}}\rangle
            +\langle\widehat{\boldsymbol\Gamma},\pounds_{\widehat{\cal Z}}\ln f_0\rangle
            \big)
\end{align*}
%\[
%Tr( [{\widehat{\boldsymbol\gamma}},  \widehat{\cal Z}^k] \widehat{\Gamma}_k)
%=
%Tr( {\widehat{\boldsymbol\gamma}}  \widehat{\cal Z}^k \widehat{\Gamma}_k
%-
%{\widehat{\boldsymbol\gamma}}\widehat{\Gamma}_k  \widehat{\cal Z}^k )
%=
%\langle {\widehat{\boldsymbol\gamma}}^\dagger,[  \widehat{\cal Z}^k ,\widehat{\Gamma}_k]
%\rangle
%\]
where $\widehat{\boldsymbol\Gamma}=i 
[\hat\rho_0
    ,            
            \nabla^{\widehat{{\gamma}}}\hat\rho_0
            ]$ and we used  the covariant Leibniz rule $\nabla\langle\widehat{A},\widehat{B}\rangle=\langle\nabla^{\widehat{{\gamma}}}\widehat{A},\widehat{B}\rangle+\langle\widehat{A},\nabla^{\widehat{{\gamma}}}\widehat{B}\rangle$ as well as the relation $[\partial_k^{\widehat{{\gamma}}},\partial_h^{\widehat{{\gamma}}}]=0$, which follows from the fact that  $\widehat{\boldsymbol{\gamma}}$ is a flat connection. As before, the Lie derivative notation is such that, for example, $\langle               \pounds_{\widehat{\cal Z}} , \widehat{\Gamma}
                    \rangle=\langle               \pounds_{\bZ} , \widehat{\boldsymbol\Gamma}
                    \rangle\cdot\de\bz$ 
                    and $\pounds_{\widehat{\cal Z}}\ln f_0=\bZ\,\lrcorner\,{\de \ln} f_0=\bZ\cdot\nabla{\ln}f_0$.
Consequently, upon introducing the Poisson tensor \eqref{PoissonBV}, we obtain 
\begin{equation}
\boldsymbol{\cal X}=
\langle\hat\rho_0,\bZ \rangle
+
\frac{\hbar}{2}
\Big(\langle\Pi\widehat{\boldsymbol\Gamma},\pounds_{\widehat{\cal Z}}\ln f_0\rangle+
\langle\pounds_{\bZ},\Pi  \widehat{\boldsymbol\Gamma}\rangle
-
\langle(\pounds_{\bZ}\Pi+[\widehat{{\gamma}}_k,\widehat{\cal Z}^k]\Pi),\widehat{\boldsymbol\Gamma}\rangle
-
\langle[\widehat{\cal Z}^k, \widehat{\Gamma}_k],\Pi{\widehat{\boldsymbol\gamma}}\rangle
\Big) 
\nonumber
%\\
%=&\,
%\langle\hat\rho_0,\boldsymbol{\cal X}_{\hmH}\rangle
%+
%\frac{\hbar}{2f_0}
%\Big( 
%\big\langle\boldsymbol{\cal X}_{\hmH}\cdot\nabla^{\widehat{\gamma}},f_0\Pi\widehat{\boldsymbol\Gamma}\big\rangle
%-
%\big\langle f_0\Pi\widehat{\boldsymbol\Gamma}\cdot\nabla^{\widehat{\gamma}},\boldsymbol{\cal X}_{\hmH} \big\rangle
%\Big)
%\label{FinalKoopVF}
\end{equation}
where $\Pi  \widehat{\Gamma} =i[\hat\rho_0
,            
\Pi\de^{\widehat{{\gamma}}}{\hat\rho_0}
]=:i[\hat\rho_0
,            
{\cal X}_{\hat\rho_0}
]$ and we recall that   $\Pi=-\Omega^{-1}$ is a Poisson \emph{bivector}, that is a skew-symmetric contravariant two-tensor. We have  
$\langle               \pounds_{\boldsymbol{\widehat{\cal W}}_1} , \boldsymbol{\widehat{\cal W}}_2
\rangle
={
\langle\widehat{{\cal W}}_1^k,\partial_k\boldsymbol{\widehat{{\cal W}}}_2\rangle
-
\langle\widehat{{\cal W}}_2^k, \partial_k\boldsymbol{\widehat{\cal W}}_1\rangle}
$, for any two operator-valued vector fields ${{\widehat{\cal W}}_1}$ and $ {\widehat{\cal W}}_2$, and 
 $             \pounds_{\boldsymbol{\widehat{\cal W}}}T =
  {\widehat{{\cal W}}}^k\partial_kT^{ij}-T^{ik}\partial_k\widehat{{\cal W}}^j- \partial_k\widehat{{\cal W}}^iT^{kj} $,  for   any contravariant two-tensor $T$.
Upon replacing $\bZ=\boldsymbol{\cal X}_{\hmH}$, we observe that
\begin{equation}
0=   \big\langle   \eta^*\big({\cal U}^\dagger (\pounds_{\bX_{\widehat{H}}} \Pi) {\cal U}\big)    , \widehat{\boldsymbol\Gamma}
    \big\rangle=
 \big\langle              (\pounds_{\boldsymbol{\cal X}_{\hmH}} \Pi), \widehat{\boldsymbol\Gamma}
    \big\rangle
-\big\langle[{\cal X}_{\hmH}^k, \widehat{\Gamma}_k],\Pi{\widehat{\boldsymbol\gamma}}\big\rangle
-\big\langle[ \widehat{\gamma}_k,\Pi \widehat{\Gamma}^k],\boldsymbol{\cal X}_{\hmH}
    \big\rangle,
\label{OmegaCons}
\end{equation}
and 
%%%%%%%%%%%%%%%%%%%%%%%%%%%
therefore we obtain
\begin{align}
    {{\cal X}}=&\,\langle
                        {\cal X}_{\hmH} \rangle
                        +\frac\hbar{2f_0}\big\langle\pounds_{{\cal X}_{\hmH}}^{\widehat{\gamma}} ,\widehat{\Xi}\big\rangle
,\qquad\qquad
\widehat{\Xi}:=i f_0[\hat\rho_0,{\cal X}_{\hat\rho_0}],
\label{FinalKoopVF}
    \end{align}
which is the convective counterpart of \eqref{HybEq2}. Indeed, the latter can be rearranged as ${X}=\langle
                        {X}_{\widehat{H}} \rangle
                        +\hbar f^{-1}\langle\pounds_{X_{\widehat{H}}},\widehat{\Sigma}\rangle/2$. Here, $\widehat{\Xi}$ is an operator-valued vector field-density and we have extended the notation for the covariant Lie derivative so that $\langle\pounds_{\boldsymbol{\cal X}_{\hmH}}^{\widehat{\gamma}} ,\widehat{\boldsymbol\Xi}\rangle
=
\langle{\cal X}_{\hmH}^k,\partial_k^{\widehat{\gamma}}\widehat{\boldsymbol\Xi}\rangle
-
\langle\widehat{\Xi}^k, \partial_k^{\widehat{\gamma}}\boldsymbol{\cal X}_{\hmH}\rangle+
\langle\widehat{\boldsymbol\Xi},\partial_k^{\widehat{\gamma}}{\cal X}_{\hmH}^k\rangle
$, where the last term vanishes due to \eqref{zerodiv}. Notice that, in this notation, equation \eqref{OmegaCons} amounts to $\pounds_{{\cal X}_{\hmH}}^{\widehat{\gamma}}\Pi=0$. The vector field  \eqref{FinalKoopVF} can also be given the following coordinate expressions:
\begin{align}\nonumber
    {\boldsymbol{\cal X}}=&\,\langle
            \hat\rho_0
                    ,
                        \boldsymbol{\cal X}_{\hmH} \rangle
                        +\frac\hbar{2f_0}\operatorname{Tr}\!\big( 
\boldsymbol{\cal X}_{\hmH}\cdot\nabla^{\widehat{\gamma}} \widehat{\boldsymbol\Xi}
-
\widehat{\boldsymbol\Xi}\cdot\nabla^{\widehat{\gamma}}\boldsymbol{\cal X}_{\hmH}
\big)
\\=&\,
\left\langle \Big(\hat\rho_0+\frac\hbar{2f_0}\operatorname{div}^{\widehat{\gamma}}\widehat{\boldsymbol\Xi}\Big),\boldsymbol{\cal X}_{\hmH}\right\rangle
+\frac\hbar{2f_0}
\operatorname{div}\operatorname{Tr}\!\big(\boldsymbol{\cal X}_{\hmH}\wedge\widehat{\boldsymbol\Xi}\big).
\label{FinalKoopVFexp}
    \end{align}
where $(\boldsymbol{\cal X}_{\hmH}\wedge\widehat{\boldsymbol\Xi})^{jk}={\cal X}_{\hmH}^j\widehat{\Xi}^k-\widehat{\Xi}^j{\cal X}_{\hmH}^k$ is a bivector density. Also, here the second equality follows from \eqref{FinalKoopVF} upon  using $\pounds_{\boldsymbol{\cal X}_{\hmH}}^{\widehat{\gamma}}\Pi=0$ and
\begin{align*}
\langle\pounds_{\boldsymbol{\cal X}_{\hmH}}^{\widehat{\gamma}} ,f_0\widetilde{\boldsymbol\Xi}\rangle
=&\,
\langle\pounds_{\boldsymbol{\cal X}_{\hmH}}^{\widehat{\gamma}} f_0,\widetilde{\boldsymbol\Xi}\rangle-\langle\pounds_{\widetilde{\boldsymbol\Xi}}^{\widehat{\gamma}} ,f_0\boldsymbol{\cal X}_{\hmH}\rangle+\langle\pounds_{\widetilde{\boldsymbol\Xi}}^{\widehat{\gamma}} f_0,\boldsymbol{\cal X}_{\hmH}\rangle
\\
=&\,
\langle\pounds_{\boldsymbol{\cal X}_{\hmH}}^{\widehat{\gamma}} ,f_0\widetilde{\boldsymbol\Xi}\rangle
-f_0\langle\pounds_{\boldsymbol{\cal X}_{\hmH}}^{\widehat{\gamma}} ,\widetilde{\boldsymbol\Xi}\rangle
-\langle\pounds_{\widetilde{\boldsymbol\Xi}}^{\widehat{\gamma}} ,f_0\boldsymbol{\cal X}_{\hmH}\rangle+\langle\pounds_{\widetilde{\boldsymbol\Xi}}^{\widehat{\gamma}} f_0,\boldsymbol{\cal X}_{\hmH}\rangle
%\\
%=&\,
%\operatorname{div}\langle{\boldsymbol{\cal X}_{\hmH}},f_0\widetilde{\boldsymbol\Xi}\rangle
%-
%f_0\langle{\widetilde{\boldsymbol\Xi}\cdot\nabla^{\widehat{\gamma}},\boldsymbol{\cal X}_{\hmH}}\rangle
%-f_0\langle\pounds_{\boldsymbol{\cal X}_{\hmH}}^{\widehat{\gamma}} ,\widetilde{\boldsymbol\Xi}\rangle
%\\
%&\,
%-\operatorname{div}\langle f_0\widetilde{\boldsymbol\Xi},{\boldsymbol{\cal X}_{\hmH}}\rangle
%+
%f_0\langle\boldsymbol{\cal X}_{\hmH}\cdot\nabla^{\widehat{\gamma}},\widetilde{\boldsymbol\Xi}\rangle
%+\langle\operatorname{div}^{\widehat{\gamma}} (f_0{\widetilde{\boldsymbol\Xi}}),\boldsymbol{\cal X}_{\hmH}\rangle
,
\end{align*}
where we have denoted  $\widetilde{\boldsymbol\Xi}=\widehat{\boldsymbol\Xi}/f_0$. Both Eulerian variants of  \eqref{FinalKoopVFexp} appeared previously in \cite{TrGB23}.

In conclusion, one obtains the following evolution equations
\begin{align}\nonumber
\frac{\partial \Omega}{\partial t}=&\ 
\big\langle\de^{\widehat{{\gamma}}}\hat\rho_0,\wedge\de^{\widehat{{\gamma}}}\hmH\big\rangle+\frac\hbar{2}\Big(\big\langle\pounds_{{\cal X}_{\hmH}}^{\widehat{\gamma}},\de^{\widehat{{\gamma}}}\widehat{\Gamma}\big\rangle-\de^{\widehat{\gamma}}\big\langle\widehat{\Gamma},\pounds_{{\cal X}_{\hmH}}\ln f_0\big\rangle\Big)
%\de\Big(\big\langle\de^{\widehat{{\gamma}}}\hmH\big\rangle+\frac\hbar{2f_0}\big\langle\pounds_{{\cal X}_{\hmH}}^{\widehat{\gamma}},f_0\widehat{\Gamma}\big\rangle\Big)
\\
\frac{\partial \hmH}{\partial t}
%=&\ 
%\Pi\Big(\de^{\widehat{{\gamma}}}\widehat{\mathcal{H}},{\big\langle
%        \de^{\widehat{{\gamma}}}\widehat{\mathcal{H}}
%    \big\rangle}+\frac\hbar{2f_0}\big\langle\pounds_{\boldsymbol{\cal X}_{\hmH}}^{\widehat{\gamma}},f_0\widehat{\boldsymbol\Gamma}\big\rangle
%\Big)
%\\
%&\,
%-
%\big[\{\hat\rho_0,
%        \widehat{\mathcal{H}}\}^{\widehat{{\gamma}}}+
%        \{\widehat{\mathcal{H}},\hat\rho_0\}^{\widehat{{\gamma}}}
%+\big[
%                \hat\rho_0
%                ,
%                 \{\ln\sqrt{f_0},
%        \widehat{\mathcal{H}}\}^{\widehat{{\gamma}}}
%            \big]
%            ,\hmH\big],
%\\
%&\,
%-\frac1{\sqrt{f_0}}
%\big[\big[\{\sqrt{f_0}\hat\rho_0,
%        \widehat{\mathcal{H}}\}^{\widehat{{\gamma}}}-\hat\rho_0\{\sqrt{f_0},
%        \widehat{\mathcal{H}}\}^{\widehat{{\gamma}}}+
%        \{\widehat{\mathcal{H}},\sqrt{f_0}\hat\rho_0\}^{\widehat{{\gamma}}}
%        +\{\sqrt{f_0},\widehat{\mathcal{H}}\}^{\widehat{{\gamma}}}\hat\rho_0
%+\big[
%                \hat\rho_0
%                ,
%                 \{\sqrt{f_0},
%        \widehat{\mathcal{H}}\}^{\widehat{{\gamma}}}
%            \big]
%            ,\hmH\big],
%\\
=&\ 
\Pi\Big(\de^{\widehat{{\gamma}}}\widehat{\mathcal{H}},{\big\langle
        \de^{\widehat{{\gamma}}}\widehat{\mathcal{H}}
    \big\rangle}+\frac\hbar{2f_0}\big\langle\pounds_{{\cal X}_{\hmH}}^{\widehat{\gamma}},f_0\widehat{\Gamma}\big\rangle
\Big)
-
\Big[\big[{
                \de^{\widehat{{\gamma}}}\hat\rho_0
+
                \hat\rho_0{\de \ln}\sqrt{{f_0}}}
                ,
                 {\cal X}_{\widehat{\cal H}}
            \big]
            ,\hmH\Big],
\label{HamKoopEvol}
    \\
i\hbar\frac{\partial \widehat{{\gamma}}}{\partial t}=&\
\de^{\widehat{{\gamma}}}\Big(\hmH+\frac{i\hbar}{2}
\big[
                2\de^{\widehat{{\gamma}}}\hat\rho_0
+
                \hat\rho_0\de \ln{f_0}
                ,
                 {\cal X}_{\widehat{\cal H}}
            \big]
\Big),
\label{gammaKoopEvol}
\end{align}
where {$\de^{\widehat{\gamma}}\widehat{\Gamma}=i[\de^{\widehat{\gamma}}\hat\rho_0,\wedge\de^{\widehat{\gamma}}\hat\rho_0]$ and}, in coordinates, $\langle\pounds_{\boldsymbol{\cal X}_{\hmH}}^{\widehat{\gamma}},f_0\widehat{\boldsymbol\Gamma}\rangle=\langle\partial_k^{\widehat{\gamma}}{{\cal X}_{\hmH}}^k,f_0\widehat{\boldsymbol\Gamma}\rangle+\langle{{\cal X}_{\hmH}}^k,\partial_k^{\widehat{\gamma}}f_0\widehat{\boldsymbol\Gamma}\rangle+\langle f_0\widehat{\Gamma}_k,\nabla^{\widehat{\gamma}}{{\cal X}_{\hmH}}^k\rangle$.
%{We have also introduced the notation $\{\widehat{A},\widehat{B}\}^{\widehat{{\gamma}}}=\Pi(\de^{\widehat{{\gamma}}}\widehat{A},\de^{\widehat{{\gamma}}}\widehat{B})$, so that \eqref{infgen2} reads
%$\hat\zeta=-\boldsymbol{\cal X}\cdot\widehat{\boldsymbol{\gamma}}-i\hbar^{-1}\hmH
%+(\{\sqrt{f_0}\hat\rho_0,{\widehat{\cal H}}\}^{\widehat{{\gamma}}}+\{{\widehat{\cal H}},\sqrt{f_0}\hat\rho_0\}^{\widehat{{\gamma}}})/{\sqrt{f_0}}$.}
Once again,  we also recall the definition  $\Pi=-\Omega^{-1}$ as in \eqref{PoissonBV}.

 The formidable appearance of these equations reflects the intricate nature of the quantum backreaction, which is responsible for the levels of complexity already appeared in the Schr\"odinger picture; see equations \eqref{HybEq1}-\eqref{HybEq3}. Nevertheless, it is possible to see that, similarly to the case of Ehrenfest dynamics, the purely quantum motion is  obtained if $\de^{\widehat{{\gamma}}}\hmH=0$, so that ${\cal X}_{\hmH}=0$ and the entire system returns trivial dynamics. Similarly, the purely classical motion is again obtained in  the case $\hmH={\cal H}\boldsymbol{1}$, so that equation \eqref{gammaKoopEvol} becomes $i\hbar{\partial_t \widehat{{\gamma}}}=\de{\cal H}$, which enforces $\widehat{{\gamma}}=i\de\phi\boldsymbol{1}$ for some function $\phi$ such that $\de\phi|_{t=0}=0$. In this case, $\de^{\widehat{{\gamma}}}=\de$ and the first two equations again return trivial dynamics.

{We also emphasize that, although in this case the local expectation $\langle\widehat{\cal H}\rangle$ is not conserved, conservation of the total energy in \eqref{KoopHam} is guaranteed by the Hamiltonian structure of the Koopman model. In the Heisenberg picture, this total energy is expressed as $h(\widehat{\mathcal{H}},\widehat{\gamma})={\int}f_0\langle \widehat{\cal H}+
   i\hbar[
                    \nabla^{\widehat{\gamma}}\hat\rho_0,        \bbX_{\hmH}
                                ]/2
    \rangle
    \de^2z_0$. As mentioned previously, here the second term  is responsible for  generating correlations between the quantum and the classical system in interaction.
}

\subsection{Heisenberg dynamics and expectation values}
\label{sec:HeisembergFull}

Following the discussion at the end of Section \ref{sec:Ehrenfest}, we may write the Heisenberg Hamitonian as a function of a set of observables, that is $\widehat{\cal H}=\widehat{H}(\widehat{{\cal O}})$ with $\widehat{{\cal O}}=({\cal U}^\dagger\widehat{{O}}_1{\cal U},{\cal U}^\dagger\widehat{{O}}_2{\cal U},\dots)\circ\boldsymbol\eta$. Then, by using \eqref{infgen2} and \eqref{FinalKoopVF}, the resulting evolution equation $\partial_t\widehat{{\cal O}}+[\hat\zeta,\widehat{{\cal O}}]=\boldsymbol{\cal X}\cdot\nabla\widehat{\cal O}$ reads explicitly as 
\begin{align*}
\frac{\partial \widehat{\cal O}}{\partial t}=&\ 
\Pi\Big(\de^{\widehat{{\gamma}}}\widehat{\cal O},{\big\langle
        \de^{\widehat{{\gamma}}}\widehat{\mathcal{H}}
    \big\rangle}
    +
    \frac\hbar{2f_0}\big\langle\pounds_{{\cal X}_{\hmH}}^{\widehat{\gamma}},f_0\widehat{\Gamma}\big\rangle
\Big)
+
\Big[i\hbar^{-1}\hmH
-\frac12
\big[
                2\de^{\widehat{{\gamma}}}\hat\rho_0
+
                \hat\rho_0{\de \ln}{f_0}
                ,
                 {\cal X}_{\widehat{\cal H}}
            \big]
            ,\widehat{\cal O}\Big],
\end{align*}
so that, {switching back to coordinate notation}, the local expectation value obeys
\begin{align}\nonumber
\frac{\partial \langle\widehat{\cal O}\rangle}{\partial t}=&\ 
%\Pi(\langle\de^{\widehat{{\gamma}}} \widehat{{\cal O}}\rangle,{\langle
%        \de^{\widehat{{\gamma}}}\widehat{\mathcal{H}}
%    \rangle})
%-
%    \frac\hbar{2f_0}\left( 
%\big\langle \widehat{\boldsymbol\Xi}\cdot,\pounds_{\boldsymbol{\cal X}_{\hmH}}^{\widehat{\gamma}}\langle\de^{\widehat{{\gamma}}}{\widehat{\cal O}}\rangle\big\rangle-\operatorname{div}\big\langle \boldsymbol{\cal X}_{\hmH},\langle\de^{\widehat{{\gamma}}}{\widehat{\cal O}}\rangle\cdot\widehat{\boldsymbol\Xi}\big\rangle
%\right)
\left\langle \boldsymbol{\cal X}_{\hmH}\cdot\langle\nabla^{\widehat{{\gamma}}} \widehat{{\cal O}}\rangle,\hat\rho_0+\frac\hbar{2f_0}\operatorname{div}^{\widehat{\gamma}}\widehat{\boldsymbol\Xi}\right\rangle
+\frac\hbar{2f_0}
\operatorname{div}{\operatorname{Tr}\!\big(\boldsymbol{\cal X}_{\hmH}\wedge\widehat{\boldsymbol\Xi}\big)}
%\!\big(\langle\boldsymbol{\cal X}_{\hmH},\widehat{\boldsymbol\Xi}\rangle-\langle\widehat{\boldsymbol\Xi},\boldsymbol{\cal X}_{\hmH}\rangle\big
\cdot\langle\nabla^{\widehat{{\gamma}}} \widehat{{\cal O}}\rangle
\\&
+\bigg\langle i\hbar^{-1}[\hmH,\widehat{{\cal O}}],\hat\rho_0+\frac\hbar{2f_0}\operatorname{div}^{\widehat{{\gamma}}}\widehat{\boldsymbol\Xi}\bigg\rangle
-
\frac1{2f_0}\big\langle 
\operatorname{div}^{\widehat{{\gamma}}}
\!\big(i\big[\widehat{\boldsymbol{\Xi}},{\hmH}\big]+f_0\big[\hat\rho_0,\big[\hat\rho_0,\boldsymbol{\cal X}_{\hmH}\big]\big]\big),\widehat{\cal O}\big\rangle,
\label{EhrKoopDyn}
\end{align}
where  have used the second line in \eqref{FinalKoopVFexp}.  The second line of \eqref{EhrKoopDyn} is obtained by repeated use of the product rule after conveniently rewriting $
                2\nabla^{\widehat{{\gamma}}}\hat\rho_0
+
                \hat\rho_0\nabla \ln{f_0}
={f_0}^{-1}\nabla^{\widehat{{\gamma}}}\widehat{P}_0+\nabla^{\widehat{{\gamma}}}({f_0}^{-1}\widehat{P}_0)$, with $\widehat{P}_0=f_0\hat\rho_0$, and  rearranging
\begin{multline*}
-\frac1{f_0}
\big\langle\widehat{P}_0,\big[\big[{f_0}^{-1}\nabla^{\widehat{{\gamma}}}\widehat{P}_0+\nabla^{\widehat{{\gamma}}}({f_0}^{-1}\widehat{P}_0),\boldsymbol{\cal X}_{\hmH}\big]
            ,\widehat{\cal O}\big]\big \rangle
%=&\,
%\frac1{f_0}\big\langle\big[\big[{f_0}^{-1}\de^{\widehat{{\gamma}}}\widehat{P}_0+\de^{\widehat{{\gamma}}}(f_0^{-1}\widehat{P}_0),\boldsymbol{\cal X}_{\hmH}\big],\widehat{P}_0\big],\widehat{\cal O}\big\rangle
\\
=
\frac1{f_0}\big\langle
\big[\boldsymbol{\cal X}_{\hmH},{f_0}^{-1}\big[\widehat{P}_0,\nabla^{\widehat{{\gamma}}}\widehat{P}_0\big]\big]+\operatorname{div}^{\widehat{{\gamma}}}\!\big[f_0^{-1}\widehat{P}_0,\big[\boldsymbol{\cal X}_{\hmH},\widehat{P}_0\big]\big]
,\widehat{\cal O}\big\rangle
%\\
%=&\,
%-\frac1{f_0}\big\langle
%\big[\de^{\widehat{{\gamma}}}{\hmH},{f_0}^{-1}\big[\widehat{P}_0,\boldsymbol{\cal X}_{\widehat{P}_0}\big]\big]
%+
%\operatorname{div}^{\widehat{{\gamma}}}\big[f_0^{-1}\widehat{P}_0,\big[\widehat{P}_0,\boldsymbol{\cal X}_{\hmH}\big]\big]
%,\widehat{\cal O}\big\rangle
%\\
%=&\,
%\frac1{f_0}\big\langle i\big[
%\operatorname{div}^{\widehat{{\gamma}}}\widehat{\boldsymbol{\Xi}},{\hmH}\big]
%-
%\operatorname{div}^{\widehat{{\gamma}}}
%\!\big(i\big[\widehat{\boldsymbol{\Xi}},{\hmH}\big]+\big[f_0^{-1}\widehat{P}_0,\big[\widehat{P}_0,\boldsymbol{\cal X}_{\hmH}\big]\big]\big),\widehat{\cal O}\big\rangle
.
\end{multline*}
This step follows by applying the Jacobi identity to the first term before the equality. 

 At this point, it may be revealing to consider the dynamics of the overall  expectation value $\int\!f_0\langle\widehat{\cal O}\rangle\,\de^2z_0$ for a purely classical and quantum observable, respectively ${\cal O}_{\sf C}$ and $\widehat{\cal O}_{\sf Q}$. Upon introducing the notation $\{\widehat{A},\widehat{B}\}^{\widehat{{\gamma}}}=\Pi(\de^{\widehat{{\gamma}}}\widehat{A},\de^{\widehat{{\gamma}}}\widehat{B})$, we have 
\begin{align*}
\frac{\partial }{\partial t}\int\!f_0{\cal O}_{\sf C}\,\de^2z_0=&\int\!
{\Big\langle\{{\cal O}_{\sf C},\widehat{\mathcal{H}}\}^{\widehat{{\gamma}}}, f_0\hat\rho_0+\frac\hbar2\operatorname{div}^{\widehat{{\gamma}}}\widehat{\boldsymbol\Xi}
    \Big\rangle}\,\de^2z_0,
\\
\frac{\partial }{\partial t}\int\!f_0\langle\widehat{\cal O}_{\sf Q}\rangle\,\de^2z_0=&\int\!
\Big\langle i\hbar^{-1}[\hmH,\widehat{{\cal O}}_{\sf Q}],f_0\hat\rho_0+\frac\hbar2\operatorname{div}^{\widehat{{\gamma}}}\widehat{\boldsymbol\Xi}\Big\rangle\,\de^2z_0,
\end{align*}
where  we have used $\langle{\cal O}_{\sf C}\rangle={\cal O}_{\sf C}$ and $\de^{\widehat{{\gamma}}}\widehat{\cal O}_{\sf Q}=0$, so that $\langle \widehat{\cal O}_{\sf Q},
\operatorname{div}^{\widehat{{\gamma}}}
\widehat{\boldsymbol{\cal W}}\rangle=\operatorname{div}\langle \widehat{\cal O}_{\sf Q},
\widehat{\boldsymbol{\cal W}}\rangle$. Also, the first equation above follows by integration by parts from the skew-symmetry of the bivector-density $\operatorname{Tr}\!\big(\boldsymbol{\cal X}_{\hmH}\wedge\widehat{\boldsymbol\Xi}\big)$. Notice that classical and quantum conservation laws follow immediately in the case $\{{\cal O}_{\sf C},\widehat{\mathcal{H}}\}^{\widehat{{\gamma}}}=0$ and $[\hmH,\widehat{{\cal O}}_{\sf Q}]=0$, respectively.
Importantly, we observe the distinctive contribution of the backreaction forces, encoded by the divergence terms, in both evolutions laws. {These terms indicate substantial differences between the Ehrenfest and Koopman models. For example, consider the translation invariant Hamiltonian $\widehat{\mathcal{H}}=(M^{-1}{\mathcal{P}}^2+m^{-1}\widehat{\!\mathscr{K}\,}\!^2)/2+V(\mathcal{Q}-\widehat{\!\mathscr{X}})$  at the end of Remark \ref{rem:ConsLaws}: as discussed, the Ehrenfest dynamics conserves the local expectation $\langle\widehat{\mathcal{M}}\rangle={\mathcal{P}}+\langle\,\widehat{\!\mathscr{K}}\rangle$, while Koopman dynamics only conserves the global momentum $\int\!f_0({\mathcal{P}}+\langle\,\widehat{\!\mathscr{K}}\rangle)\,\de^2z_0$. 
Indeed, since $\de^{\widehat{{\gamma}}\,}\widehat{\!\mathscr{X}}=0$, we have $\de^{\widehat{{\gamma}}}V=\partial_{\mathcal{Q}}V\, \de\mathcal{Q}$ and $\{\mathcal{P},\widehat{\mathcal{H}}\}^{\widehat{{\gamma}}}=i\hbar^{-1}[\,\widehat{\!\mathscr{K}},\hmH]$.
We notice that the total  momentum conservation may be related to Galilean invariance in the hybrid quantum-classical setting \cite{Bermudez}, although here we leave this aspect for future work.}

While the expectation value dynamics offers some insight, the intricate structure of the terms associated to the quantum backreaction leaves little room to physical intuition. At present, the role of these terms in mixed quantum-classical Heisenberg dynamics  can hardly be understood without resorting to suitable examples. In particular, here we want to compare the Ehrenfest and the Koopman models for a specific problem.

\section{Mixed quantum-classical pure-dephasing systems}
\label{sec:pureDephasing}

In this section, we present the Heisenberg dynamics for the case of pure-dephasing systems. In particular, we consider the   Hamiltonian
\beq
\widehat{H}=\frac12(p^2+q^2)+\alpha q\widehat{\sigma}_z\implies\hmH=\frac12({\cal P}^2+{\cal Q}^2)+\alpha {\cal Q}\widehat{\Sigma}_z=:{\cal H}_o+\alpha {\cal Q}\widehat{\Sigma}_z
\label{PDHam}
\eeq
where  $({\cal Q},{\cal P})=(\eta^*q,\eta^*p)$ and $\widehat{\Sigma}_z=\eta^*({\cal U}^\dagger\widehat{\sigma}_z{\cal U})$. Here, we have set the oscillator constants to 1 for simplicity. This type of Hamiltonian emerges as the simplest type of Jahn-Teller system in chemical physics and its quantum-classical dynamics was recently considered  in \cite{Manfredi23}. For later purpose, we compute the covariant differential, that is
\[
\de^{\widehat{\gamma}}\widehat{\cal H}=\frac12\de({\cal P}^2+{\cal Q}^2)+\alpha\widehat{\Sigma}_z\de{\cal Q},
\]
where we recall that  $0=\eta^*({\cal U}^\dagger\de\widehat{\sigma}_z{\cal U})=\de^{\widehat{\gamma}}\widehat{\Sigma}_z$. {In this section, we will mostly focus on the Koopman formulation and restrict to consider the Ehrenfest case depending on convenience}.

We start by looking at the dynamics in the purely quantum sector. The Heisenberg dynamics of the vertical component of the spin reads
\begin{align*}
\frac{\partial \widehat{\Sigma}_z}{\partial t}=&\ 
\frac\alpha{\sqrt{f_0}}\{{\cal Q},\sqrt{f_0}[[\hat\rho_0,\widehat{\Sigma}_z],\widehat{\Sigma}_z]\}^{\widehat{{\gamma}}}
%\\
%=&\ 
%\frac{2\alpha}{{f_0}}\sqrt{f_0}\{{\cal Q},\sqrt{f_0}(\hat\rho_0-\widehat{\Sigma}_z\hat\rho_0\widehat{\Sigma}_z)\}^{\widehat{{\gamma}}}
\\
=&\ 
\frac{\alpha}{{f_0}}
\big(2\{{\cal Q},{f_0}(\hat\rho_0-\widehat{\Sigma}_z\hat\rho_0\widehat{\Sigma}_z)\}^{\widehat{{\gamma}}}
-
\{{\cal Q},{f_0}\}^{\widehat{{\gamma}}}(\hat\rho_0-\widehat{\Sigma}_z\hat\rho_0\widehat{\Sigma}_z)\big),
\end{align*}
which is generally not conserved. We observe that the nonconservative terms are triggered exclusively by the backreaction contribuition from the Koopman model, which are instead absent in the Ehrenfest dynamics. Upon noticing $\widehat{\Sigma}_z^2=\widehat{\sigma}_z^2=\boldsymbol{1}$ and setting $\hat\rho_0=\psi_0\psi_0^\dagger$, so that $\langle\hat\rho_0,\de^{\widehat{{\gamma}}}\hat\rho_0\rangle=0$, the local expectation values dynamics reads
\begin{align*}
\frac{\partial \langle\widehat{\Sigma}_z\rangle}{\partial t}
=&\,
\frac{\alpha}{{f_0}}
\big(2\{{\cal Q},{f_0}(1-\langle\widehat{\Sigma}_z\rangle^2 )\}^{\widehat{{\gamma}}}
-
2{f_0}\langle\{{\cal Q},\hat\rho_0\}^{\widehat{{\gamma}}},\hat\rho_0-\widehat{\Sigma}_z\hat\rho_0\widehat{\Sigma}_z\rangle
-
\{{\cal Q},{f_0}\}^{\widehat{{\gamma}}}(1-\langle\widehat{\Sigma}_z\rangle^2)  \big)
\\
=&\,
\frac{\alpha}{{f_0}}
\big(2\{{\cal Q},{f_0}(1-\langle\widehat{\Sigma}_z\rangle^2)\}^{\widehat{{\gamma}}}
+
{f_0}\{{\cal Q},\langle\widehat{\Sigma}_z\rangle^2\}^{\widehat{{\gamma}}}
-
\{{\cal Q},{f_0}\}^{\widehat{{\gamma}}}(1-\langle\widehat{\Sigma}_z\rangle^2)\big)
\\
=
&\, 
\frac\alpha{f_0}
\{{\cal Q},f_0(1-\langle\widehat{\Sigma}_z\rangle^2)\}^{\widehat{{\gamma}}}
\end{align*}
so that $\int\! f_0\langle\widehat{\Sigma}_z\rangle\,\de^2z_0=const.$ Thus, we observe that, while the local expectation is generally not conserved, the conservation law holds instead for overall expectation value.

We now move on to consider the Heisenberg dynamics in the classical sector.
We evaluate
\[
{\cal X}_{\hmH}=\Pi\de{{\cal H}_o}+\alpha\Pi\de{\cal Q}\widehat{\Sigma}_z={\cal X}_{{\cal H}_o}+\alpha\widehat{\Sigma}_z{\cal X}_{\cal Q}
\]
so that,
\begin{align*}
\big\langle\pounds_{{\cal X}_{\hmH}}^{\widehat{\gamma}},f_0\widehat{\Gamma}\big\rangle
%=&\,
%\big\langle{\cal X}_{\cal Q}^k\widehat{\Sigma}_z,\partial_k^{\widehat{\gamma}}(f_0\widehat{\boldsymbol\Gamma})\big\rangle
%+
%f_0\big\langle\widehat{\Gamma}_k,\nabla^{\widehat{\gamma}}{\cal X}_{\cal Q}^k\widehat{\Sigma}_z\big\rangle
%\\
%=&\,
%\operatorname{Tr}\big(\partial_k^{\widehat{\gamma}}({\cal X}_{\cal Q}^k\widehat{\Sigma}_zf_0\widehat{\boldsymbol\Gamma})\big)
%+
%f_0\big\langle\widehat{\Gamma}_k,\widehat{\Sigma}_z\big\rangle\nabla{\cal X}_{\cal Q}^k
%\\
%=&\,
%\partial_k\big({\cal X}_{\cal Q}^k\langle\widehat{\Sigma}_z,f_0\widehat{\boldsymbol\Gamma}\rangle\big)
%+
%f_0\big\langle\widehat{\Gamma}_k,\widehat{\Sigma}_z\big\rangle\nabla{\cal X}_{\cal Q}^k
%\\
=&\,
\pounds_{{\cal X}_{\cal Q}}\big\langle\widehat{\Sigma}_z,f_0\widehat{\Gamma}\big\rangle
=
\pounds_{{\cal X}_{\cal Q}}\big\langle if_0\de^{\widehat{\gamma}}[\hat\rho_0,\widehat{\Sigma}_z]\big\rangle
\end{align*}
and  equation \eqref{EhrKoopDyn} leads to
\[
\frac{\partial {\cal Q}}{\partial t}= {\cal P}
-
    \frac{\alpha\hbar}{2f_0}
\big\{{\cal Q},f_0\big\langle i\{{\cal Q},[\hat\rho_0,\widehat{\Sigma}_z]\}^{\widehat{\gamma}}\big\rangle\big\}^{\widehat{\gamma}}
\,,\qquad\qquad
\frac{\partial {\cal P}}{\partial t}= -{\cal Q}
-
    \frac{\alpha\hbar}{2f_0}
\big\{{\cal P},f_0\big\langle i\{{\cal Q},[\hat\rho_0,\widehat{\Sigma}_z]\}^{\widehat{\gamma}}\big\rangle\big\}^{\widehat{\gamma}},
\]
so that the overall expectation values $\int\! f_0{\cal Q}\,\de^2z_0$ and $\int\! f_0{\cal P}\,\de^2z_0$ undergo simple harmonic motion. {This follows from the fact that $\{{\cal Q},f_0\langle i\{{\cal Q},[\hat\rho_0,\widehat{\Sigma}_z]\}^{\widehat{\gamma}}\rangle\}^{\widehat{\gamma}}=\Pi(\de{\cal Q},\de \langle if_0\{{\cal Q},[\hat\rho_0,\widehat{\Sigma}_z]\}^{\widehat{\gamma}}\rangle)=-\operatorname{div}(\langle if_0\{{\cal Q},[\hat\rho_0,\widehat{\Sigma}_z]\}^{\widehat{\gamma}}\rangle{\boldsymbol{\cal X}}_{\cal Q})$, and analogously for the corresponding term in the momentum equation.    We notice that this expectation dynamics again differs from the underlying Heisenberg evolution of ${\cal Q}$ and ${\cal P}$}, which instead carries extra terms associated to the backreaction contributions. In the case of the Ehrenfest model, these contributions are absent and one is left with simple harmonic motion also at the level of the underlying Heisenberg dynamics.

Notice that the equations above must also be accompanied by 
\begin{align*}
\frac{\partial \Omega}{\partial t}=&\ \alpha\de{\cal Q}\wedge\de\langle\widehat{\Sigma}_z\rangle
%+
%\frac\hbar{2}\big\langle\widehat{\Sigma}_z\pounds_{\boldsymbol{\cal X}_{\cal Q}}^{\widehat{\gamma}},\de^{\widehat{\gamma}}\widehat{\boldsymbol\Gamma}\big\rangle
+\alpha
\frac{\hbar}{2}\Big(\big\langle\widehat{\Sigma}_z,\de^{\widehat{\gamma}}\{\widehat{\Gamma},{\cal Q}\}^\gamma\big\rangle
\\
&\,
+\de\{\ln f_0,{\cal Q}\}^{\widehat{{\gamma}}}\wedge\big\langle\widehat{\Sigma}_z,\widehat{\Gamma}\big\rangle+\{\ln f_0,{\cal Q}\}^{\widehat{{\gamma}}}\big\langle\widehat{\Sigma}_z,\de^{\widehat{\gamma}}\widehat{\Gamma}\big\rangle\Big)
    \\
i\hbar\frac{\partial \widehat{{\gamma}}}{\partial t}=&\
\de {\cal H}_o+\alpha\widehat{\Sigma}_z\de{\cal Q}+
\alpha\frac{i\hbar}{2}
\big[
                \de^{\widehat{{\gamma}}}\big(2\{\hat\rho_0,{\cal Q}\}^{\widehat{{\gamma}}}
+
                \hat\rho_0\{ \ln{f_0},{\cal Q}\}^{\widehat{{\gamma}}}\big)
                ,
                 \widehat{\Sigma}_z
            \big] 
,
\end{align*}
where the notation is such that $\{\widehat{\Gamma},{\cal Q}\}^\gamma={\cal X}_{Q}\,\lrcorner\,\de^{\widehat{\gamma}}\widehat{\Gamma}$. We observe that the Koopman model introduces terms orthogonal to $\widehat{\Sigma}_z$ in the second equation. These terms are absent in the Ehrenfest model{, for which $\partial_t\langle\widehat{\Sigma}_z\rangle=0$}. Moreover, in the latter case, setting $\langle\widehat{\Sigma}_z\rangle=0$ yields $\partial_t\Omega=0$  {so that}, as we saw above, the dynamics of the classical variables completely decouples from the quantum dynamics. The situation is very different for the Koopman model, for which {$\langle\widehat{\Sigma}_z\rangle$ is not conserved and} the quantum-classical coupling persists at all times due to the backreaction terms.

Evidently, the quantum backreaction produces extra  terms in both the velocity and the force which are  completely overlooked by the Ehrenfest model. Already appearing in the Schr\"odinger picture \cite{GBTr22a}, this crucial difference makes the Heisenberg dynamics for the Koopman model more realistic and compatible with the fully quantum description. Indeed, in the latter case, the dynamics of the orbital degrees of freedom cannot  decouple from the spin motion, thereby leading to a situation very different from the result of the Ehrenfest model and closer to Koopman quantum-classical dynamics.

%%%%%%%%%%%%%%%%%%%%%

\section{Conclusions and perspectives}
\label{sec:conclusions}

Motivated by the importance of mixed quantum-classical (MQC) models in a variety of contexts, we have presented the hybrid convective-Heisenberg representation of mixed quantum-classical dynamics. Unlike standard quantum mechanics, the mixed quantum-classical counterpart is made particularly challenging by the interplay between the Lagrangian paths advancing classical phase-space observables and the unitary operators that are responsible of quantum evolution. This interplay makes the Heisenberg representation of MQC systems quite involved and in, some cases, counterintuitive. For example, the Heisenberg Hamiltonian operator is not conserved in any of the two descriptions considered here. In this paper, we overcame the challenges posed by  quantum-classical coupling by resorting to the action principle formulation underlying both the Ehrenfest and the Koopman models. Indeed, the geometric structure underlying their variational formulation allowed us to write explicit Heisenberg equations for MQC dynamics. Importantly, the diffeomorphic Lagrangian paths on phase-space do not preserve the  symplectic structure, which then possesses its own Lie-transport equation $\partial_t\Omega=\pounds_{\cal X}\Omega$. The latter is necessary to characterize the convected Poisson bracket $\{\cdot,\cdot\}^{\widehat{\gamma}}=\Pi(\de^{\widehat{\gamma}}\cdot,\de^{\widehat{\gamma}}\cdot)$, which in turn needs the evolution equation  for the pure-gauge connection $\widehat{\gamma}$. The latter has a dominant role throughout the entire construction, which evidently necessitates covariant derivatives all along.  As we have seen, the covariant Lie derivative $\pounds^{\widehat\gamma}$ also plays a major role, especially in the Koopman model. Despite this rich geometric structure, MQC observables do not appear to have a Lie algebra structure, thereby confirming the results obtained in previous studies.

Of the two descriptions considered here, the Ehrenfest model is certainly simpler. While this model is known to fail in reproducing important correlation effects, its underlying geometric structure is particularly rich and has set up the ground for the study of the Koopman model. In the latter, the backreaction term comprising statistical correlations appears explicitly in the total energy, following an analogy with spin-orbit coupling in semi-relativistic mechanics. As already apparent in the Eulerian Schr\"odinger picture, this backreaction term lifts the difficulty of the equations to a formidable level. Nevertheless, we were able to prove that, for purely (classical) quantum observables, a conservation law for the overall expectation value becomes apparent whenever the observable (Poisson) commutes with the Heisenberg Hamiltonian. This applies to both the models considered here. As for the {local} expectation $\langle\widehat{
\cal H}\rangle$ of the Hamiltonian itself, we emphasize that this is conserved for the Ehrenfest dynamics, but not for the Koopman case. Evidently, in the latter case this is due to the presence of the backreaction term. This term was shown to be particularly relevant in the simple example of pure dephasing dynamics, in which case the Ehrenfest dynamics becomes trivial for the classical canonical observables and for the vertical  component of the quantum spin. {We remind the reader that the local expectation $\langle\widehat{
\cal H}\rangle$ must not be confused with the total energy ${\int}f_0\langle\widehat{
\cal H}\rangle\de^2z_0$ which is consistently conserved in both the Ehrenfest model. Likewise, the Koopman model conserves the energy functional ${\int}f_0\langle \widehat{\cal H}+
   i\hbar[
                    \nabla^{\widehat{\gamma}}\hat\rho_0,        \bbX_{\hmH}
                                ]/2
    \rangle
    \de^2z_0$, whose second term generates quantum-classical correlations}.

Given the construction presented here, one can think of different variants depending on the requirements posed by the specific problem under consideration. For example, one can merge the Eulerian Schr\"odinger picture and the convective Heisenberg picture to obtain an MQC counterpart of Dirac's interaction picture. In addition, one may also have hybrid descriptions in which, for example, the classical degrees of freedom are treated in the Eulerian frame while the quantum observables are formulated in the Heisenberg picture. Indeed, this type of hybrid description is made possible by the semidirect-product group structure underlying the original models in  terms of diffeomorphic phase-space paths and unitary propagators in the quantum sector. These alternative descriptions are left for future work. 

{We conclude by emphasizing that, despite the level of complexity in the Heisenberg dynamics, both the Ehrenfest and Koopman models possess convenient particle closures \cite{TrGB23} that are formulated combining their underlying variational principle with the momentum map structures associated to their geometry. In the Koopman case, the numerical implementation has been successfully benchmarked in \cite{BaBeGBTr24}. This is an example in which the complexity of a PDE system may be conveniently overcome by resorting to its underlying geometric structure. In this sense, while the hybrid Schr\"odinger picture remains considerably simpler, the geometric structure of the corresponding Heisenberg dynamics presented here may lead to new directions to be unfolded in the future. One possible example is to transfer the Heisenberg picture --or any of its extensions proposed above-- to the level of the particle closure in such a way to tackle specific problems where the Schr\"odinger picture is less convenient.}

\medskip
\paragraph*{Data Availability Statement:} No Data associated in the manuscript

\paragraph{Acknowledgments.} We are grateful to Werner Bauer, Paul Bergold, Jes\'us Clemente-Gallardo, Fran\c{c}ois Gay-Balmaz, Darryl Holm, Alberto Ibort, and Delyan Zhelyazov for their comments and remarks on this and related topics. 
This work was made possible through the support of Grant 62210 from the John Templeton Foundation. The opinions expressed in this publication are those of the authors and do not necessarily reflect the views of the John Templeton Foundation. Financial support by the Leverhulme Research Project Grant RPG-2023-078 is also greatly acknowledged.  DMC also acknowledges financial support by Gobierno de Aragón through the grant defined in ORDEN CUS/581/2020 as well as the support of Grant PID2021-123251NB-I00 funded by MCIN/AEI/10.13039/ 501100011033 and by the European Union, and of Grant E48-23R funded by Government of Aragon..


\bigskip

\begin{thebibliography}{99}

\bibitem{AkLoPr14}
Akimov, A.V.; Long, R.; Prezhdo, O.V.  {\it Coherence penalty functional: A simple method for adding decoherence in Ehrenfest dynamics}. J. Chem. Phys. 140 (2014), 194107

\bibitem{Aleksandrov}
Aleksandrov, I.V.
{\it The statistical dynamics of a system consisting of a classical and a quantum subsystem}.
{Z. Naturforsch.} 36a (1981), 902-908


\bibitem{AlBoClMa24}
Alonso, J.L.; Bouthelier-Madre, C.; Clemente-Gallardo, J.; Mart\'inez-Crespo, D. {\it Hybrid geometrodynamics: a Hamiltonian description of classical gravity coupled to quantum matter}. Classical Quantum Gravity 41 (2024), 105004

  \bibitem{alonsoEffectiveNonlinear2023}
    Alonso, J.L.; Bouthelier-Madre, C.; Clemente-Gallardo, J.; Martínez-Crespo, D.; Pomar, J. {\it  Effective nonlinear Ehrenfest hybrid quantum-classical dynamics}. Eur. Phys. J. Plus 138 (2023), 649


\bibitem{Alonso}
Alonso, J.L.; Clemente-Gallardo, J.; Cuch\'i, J.C.; Echenique, P.; Falceto, F.
{\it Ehrenfest dynamics is purity non-preserving: A necessary ingredient for decoherence}.
J. Chem. Phys. 137 (2012), 054106



\bibitem{Anderson}
Anderson, A. {\it Quantum backreaction on ``classical'' variables}. {Phys. Rev. Lett.} 74 (1995), 621-625

\bibitem{BaBeGBTr24}
Bauer, W.; Bergold, P.; Gay-Balmaz, F.; Tronci, C. {\it  Koopmon trajectories in nonadiabatic quantum-classical dynamics}.  Multiscale Model. Simul.  {22 (2024), n. 4,  1365-1401}

\bibitem{Baym}
Baym, G. Lectures On Quantum Mechanics. CRC Press. Boca Raton FL, USA. 1969.

{
\bibitem{Bermudez}
Berm\'udez Manjarres, A.D. {\it  Projective representation of the Galilei group for classical and quantum-classical systems}. J. Phys. A: Math. Theor. 54 (2021), n. 44, 444001
}

\bibitem{BBirula}
Bialynicki-Birula, I.; Cieplak, M.; Karminski, J.; Furdyna, A.M. Theory of Quanta. Oxford University Press. 1992

\bibitem{BoGBTr19}
Bondar, D.I.; Gay-Balmaz, F.; Tronci, C.
{\it Koopman wavefunctions and classical-quantum correlation dynamics}.
Proc. R. Soc. A 475 (2019), n. 2229, 20180879

\bibitem{Bondarenko}
Bondarenko, A.S.; Tempelaar, R.
{\it Overcoming positivity violations for density matrices in surface hopping}.
J. Chem. Phys. 158 (2023), n. 5, 054117

\bibitem{BLTr15}
Bonet-Luz, E.; Tronci, C. {\it Geometry and symmetry of quantum and classical-quantum variational principles}. J. Math. Phys. 56 (2015), 082104

\bibitem{Traschen}
Boucher, W.; Traschen, J.
{\it Semiclassical physics and quantum fluctuations}.
Phys. Rev. D 37 (1988), 3522-3532

\bibitem{boutheliermadre2023}
Bouthelier-Madre, C. ; Clemente-Gallardo, J. ; González-Bravo, L. ; Martínez-Crespo, D. 
{\it Hybrid Koopman $C^*$-formalism and the hybrid quantum-classical master equation}.
J. Phys. A: Math. Gen. 56 (2023), 374001


\bibitem{Marmo}
Chru\'sci\'nski, D.; Kossakowski, A.; Marmo, G.; Sudarshan, E.C.G.
{\it Dynamics of interacting classical and quantum systems}.
Open. Syst. Inf. Dyn. 18 (2011), n. 4, 339-351

\bibitem{CrBa18}
Crespo-Otero, R.; Barbatti, M.
{\it Recent advances and perspectives on nonadiabatic mixed quantum-classical dynamics}.
Chem. Rev. 118 (2018), n. 15, 7026-7068

\bibitem{Diosi}
Di\'osi, L.; Halliwell, J.J.
{\it Coupling classical and quantum variables using continuous quantum measurement theory}.
Phys. Rev. Lett. 81 (1998), 2846

\bibitem{FoTr24}
Foskett, M.S.; Tronci, C. {\it Holonomy and vortex structures in quantum hydrodynamics}. 
Math. Sci. Res. Inst. Publ. 72 (2024), 173--214

\bibitem{GBMaRa12}
Gay-Balmaz, F.; Marsden, J.E.; Ratiu, T.S. {\it Reduced variational formulations in free boundary continuum mechanics}.  J. Nonlinear Sci. 22 (2012), 463-497

{
\bibitem{GBRa}
Gay-Balmaz, F.; Ratiu, T.S. {\it The geometric structure of complex fluids}.
 Adv. Applied Math.,  42 (2009), n. 2, 176-275

\bibitem{GBRaTr}
Gay-Balmaz, F.; Ratiu, T.S.; Tronci, C.
 {\it Euler-Poincar\'e approaches to nematodynamics}.
 Acta Appl. Math., 120 (2012), 127-151
 }

\bibitem{GBTr23}
Gay-Balmaz, F.; Tronci, C.
{\it Dynamics of mixed quantum-classical spin systems}.
J. Phys. A 56 (2023), 144002 

\bibitem{GBTr22a}
Gay-Balmaz, F.; Tronci, C.
{\it Evolution of hybrid quantum-classical wavefunctions}.
Phys. D 440 (2022), 133450 

\bibitem{GBTr-fluid}
Gay-Balmaz, F.; Tronci, C.
{\it Complex fluid models of mixed quantum-classical dynamics}.  J. Nonlinear Sci. {34 (2024), 81}

\bibitem{GBTr22}
Gay-Balmaz, F.; Tronci, C. {\it Koopman wavefunctions and classical states in hybrid quantum-classical dynamics}. J. Goem. Mech. 14 (2022), 559-596.

\bibitem{GBTr20}
Gay-Balmaz, F.; Tronci, C.
{\it Madelung transform and probability densities in hybrid quantum-classical dynamics}.
Nonlinearity, 33 (2019), n. 10, 5383-5424
    
\bibitem{Gerasimenko}
Gerasimenko, V.
{\it Dynamical equations of quantum-classical systems}.
Theor. Math. Phys. 50 (1982), 49-55 

\bibitem{Gerasimenko2}
Gerasimenko, V.
{\it On evolution equations of quantum-classical systems}. {\tt arXiv:0909.4942}

\bibitem{GiGrSc23}
Giulini, D.; Gro\ss ardt, A.; Schwartz, P.K. {\it Coupling quantum matter and gravity}. Lecture Notes in Phys. 1017 (2023), 491-550

\bibitem{Hall}
Hall, M.J.W.; Reginatto, M.
Ensembles on Configuration Space.
Springer. 2016

{
\bibitem{Holm02}
Holm,  D.D. {\it Euler-Poincar\'e dynamics of perfect complex fluids}, in Geometry, Mechanics,
and Dynamics. P. Holmes, P. Newton \& A. Weinstein (Eds.). Springer, New York. 2002
}

\bibitem{HoMaRa98}
Holm, D.D.; Marsden, J.E.; Ratiu, T.S.
{\it The Euler-Poincar{\'e} equations and semidirect products with applications to continuum theories}.
Adv. Math. 137 (1998), 1-81
    
\bibitem{HoMaRa86}
Holm, D.D.; Marsden, J.E.; Ratiu, T. S. {\it The Hamiltonian structure of continuum mechanics in material, inverse material, spatial and convective representations}. S\'em. Math. Sup., 100 (1986), 11-124

\bibitem{JaSu10}
Jauslin, H.R.; Sugny, D.
{\it Dynamics of mixed quantum-classical systems, geometric quantization and coherent states}. Lect. Notes Ser. Inst. Math. Sci. Natl. Univ. Singap., 20 (2010), 65-96

\bibitem{Kapral}
Kapral R.
{\it Progress in the theory of mixed quantum-classical dynamics}.
Annu. Rev. Phys. Chem. 57 (2006), 129–157.

\bibitem{Koopman}
Koopman, B.O.
{\it Hamiltonian systems and transformations in Hilbert space}.
Proc. Nat. Acad. Sci. 17 (1931), 315

\bibitem{Madelung} 
Madelung, E. {\it Quantentheorie in hydrodynamischer Form}. {Z. Phys.} 40 (1927), 322-326

\bibitem{MaHuHe19}
Manfredi, G.; Hervieux P.-A.; Hurst, J. 
{\it  Phase-space modeling of solid-state plasmas}.
Rev. Mod. Plasma Phys. 3  (2019), 13

\bibitem{Manfredi23}
Manfredi, G.; Rittaud, A.; Tronci, C.
{\it Hybrid quantum-classical dynamics of pure-dephasing systems}.
J. Phys. A: Math. Theor. 56 (2023) 154002 

\bibitem{Mauro}
Mauro, D. {\it On Koopman-von Neumann waves}. {Int. J. Mod. Phys. A} 17 (2002), 1301

\bibitem{PeTe}
Peres, A.; Terno, D.R. {\it Hybrid classical-quantum dynamics}. {Phys. Rev. A} 63 (2001), 022101 

\bibitem{PrKi}
Prezhdo, O.V.; Kisil, V.V. {\it Mixing quantum and classical mechanics}. {Phys. Rev. A} 56 (1997), 162-175

\bibitem{Sahoo}
Sahoo, D. {\it Mixing quantum and classical mechanics and uniqueness of Planck's constant}. {J. Phys. A: Math. Gen.} 37 (2004), 997-1010

%\bibitem{SeKeCiKa03}
%Sergi, A.; Kernan, D.M.; Ciccotti, G.; Kapral, R. {\it Simulating quantum dynamics in classical environments}. Theor. Chem. Acc. 110 (2003), 49-58

\bibitem{SiMaKr88}
Simo, J.C.; Marsden, J.E.; Krishnaprasad, P.S. {\it The Hamiltonian structure of nonlinear elasticity: the material and convective representations of solids, rods, and plates}.  Arch. Rational Mech. Anal. 104 (1988), 125-183.

\bibitem{Sudarshan}
Sudarshan, E.C.G.
{\it Interaction between classical and quantum systems and the measurement of quantum observables}.
{Pr\={a}ma\d{n}a} 6 (1976), n. 3, 117-126.

\bibitem{Thomas}
Thomas, L.H. {\it The kinematics of an electron with an axis}. Phil. Mag. 7 (1927), 1-23

{
\bibitem{TrMCGB}
Tronci, C.; Mart\'inez-Crespo, D.; Gay-Balmaz, F. {\it Entropy functionals and equilibrium states in mixed quantum-classical dynamics}. Lecture Notes in Comput. Sci. (submitted). {\tt arXiv:2501.18587}
}

\bibitem{TrGB23}
Tronci, C.; Gay-Balmaz, F.
{\it Lagrangian trajectories and closure models in mixed quantum-classical dynamics}.
Lecture Notes in Comput. Sci. 14072 (2023), 290-300

\bibitem{Tully90}
Tully, J.C.
{\it Molecular dynamics with electronic transitions}.
J. Chem. Phys. 93, 1061 (1990)

\bibitem{WiZe84}
Wilczek, F.; Zee, A. {\it Appearance of gauge structure in simple dynamical systems}.  Phys. Rev. Lett. 52 (1984), 2111-2114
 


    \end{thebibliography}
\end{document}